\documentclass[pdflatex,sn-nature]{sn-jnl}

\usepackage[T1]{fontenc}
\usepackage{textcomp}
\usepackage{amsmath}
\usepackage{amsfonts}
\usepackage{amssymb}
\usepackage{graphicx}
\usepackage{siunitx}
\usepackage{setspace}
\usepackage{xcolor}
\usepackage{threeparttable}
\usepackage{booktabs}
\usepackage{bm}
\usepackage{makecell}
\begin{document}

\title{Hybrid BaTiO$_3$/TiO$_2$ Metasurface for Efficient Gigahertz-Speed Free-Space Electro-Optic Modulation}

\author[1,2]{\fnm{Zhongpeng} \sur{Sun}}
\equalcont{These authors contributed equally to this work.}

\author[1]{\fnm{Kerolos M. A.} \sur{Yousef}}
\equalcont{These authors contributed equally to this work.}

\author[1]{\fnm{Michael} \sur{Domm}}

\author[3]{\fnm{Agham} \sur{Posadas}}

\author[1]{\fnm{Xudong} \sur{Li}}

\author[1,4]{\fnm{Marcus} \sur{Ossiander}}

\author[1,5]{\fnm{Maryna L.} \sur{Meretska}}

\author[6]{\fnm{Yiwei} \sur{Ju}}

\author[1]{\fnm{Isabel} \sur{Barth}}

\author[7]{\fnm{Jason} \sur{Tischler}}

\author[1]{\fnm{Theodore P.} \sur{Letsou}}

\author[6]{\fnm{Moaz} \sur{Waqar}}

\author[8]{\fnm{Amirhassan} \sur{Shams-Ansari}}

\author[6,9,10]{\fnm{Xiaoqing} \sur{Pan}}

\author[7]{\fnm{Alexander A.} \sur{Demkov}}

\author*[1]{\fnm{Federico} \sur{Capasso}}\email{capasso@seas.harvard.edu}

\affil[1]{\orgname{Harvard John A. Paulson School of Engineering and Applied Sciences, Harvard University}, 
\orgaddress{\city{Cambridge}, \state{Massachusetts}, \postcode{02138}, \country{USA}}}

\affil[2]{\orgname{Department of Chemistry and Chemical Biology, Harvard University}, 
\orgaddress{\city{Cambridge}, \state{Massachusetts}, \postcode{02138}, \country{USA}}}

\affil[3]{\orgname{La Luce Cristallina, Inc.}, 
\orgaddress{\city{Austin}, \state{Texas}, \postcode{78759}, \country{USA}}}

\affil[4]{\orgname{Institute of Experimental Physics, Graz University of Technology}, 
\orgaddress{\city{Graz}, \country{Austria}}}

\affil[5]{\orgname{Institute of Nanotechnology, Karlsruhe Institute of Technology}, 
\orgaddress{\city{Karlsruhe}, \country{Germany}}}

\affil[6]{\orgname{Department of Materials Science and Engineering, University of California, Irvine}, 
\orgaddress{\city{Irvine}, \state{California}, \postcode{92697}, \country{USA}}}

\affil[7]{\orgname{Department of Physics, The University of Texas at Austin}, 
\orgaddress{\city{Austin}, \state{Texas}, \country{USA}}}

\affil[8]{\orgname{Monarch Quantum}, 
\orgaddress{\city{San Diego}, \state{California}, \postcode{92131}, \country{USA}}}

\affil[9]{\orgname{Department of Physics and Astronomy, University of California, Irvine}, 
\orgaddress{\city{Irvine}, \state{California}, \country{USA}}}

\affil[10]{\orgname{Irvine Materials Research Institute (IMRI), University of California, Irvine}, 
\orgaddress{\city{Irvine}, \state{California}, \country{USA}}}


\abstract{%
Free-space electro-optic modulators are key to emerging photonic systems, yet their performance remains limited by trade-offs between modulation efficiency, bandwidth, and device aperture. Here we report a hybrid BaTiO$_3$ (BTO)/TiO$_2$ metasurface for large-aperture, efficient, gigahertz-speed free-space electro-optic modulation. Combining scalable BTO film growth by radio-frequency magnetron sputtering with mature TiO$_2$ nanofabrication, we pattern the metasurface in TiO$_2$ on an unetched BTO layer. The resulting devices support guided-mode resonances with quality factors exceeding 1300 and an optical confinement factor of $\sim 0.8$, while the continuous BTO layer makes efficient use of the applied voltage, together maximizing the overlap between the optical and driving fields within the BTO. A device with a $0.3~\mathrm{mm} \times 0.3~\mathrm{mm}$ metasurface achieves a transmittance modulation efficiency of $\sim 0.020~\mathrm{V^{-1}}$ and a $-3~\mathrm{dB}$ electro-optic bandwidth of $\sim 0.8~\mathrm{GHz}$, with an effective Pockels coefficient of $\sim 151~\mathrm{pm/V}$ for the BTO. This establishes a scalable route to high-performance free-space electro-optic modulators for LiDAR, free-space optical communication, and reconfigurable optical computing.
}

% \keywords{metasurface, photonics, nonlinear optics}

\maketitle

\section{Introduction}

Electro-optic (EO) systems convert electrical signals into optical responses, offering low loss, high speed, and enhanced multiplexing capabilities compared with purely electronic systems~\cite{McMahon2023Reveiw}. These features underpin a range of frontier technologies, from optical interconnects in data centers~\cite{Miller2017Interconnect} to qubit manipulation in photonic~\cite{PsiQuantum2025BTO} and neutral-atom quantum computing~\cite{Weiss2017NeuAtom}. While on-chip EO modulators have been extensively developed~\cite{Wang2018LNO,raju2025high}, free-space EO modulators remain relatively underexplored, despite their unique potential in light detection and ranging (LiDAR)~\cite{Park2021Lidar}, free-space optical communication~\cite{Zhu2021FSO}, and scalable optical computing~\cite{Hu2024OpComp}.

Metasurfaces have emerged as a promising platform for free-space EO modulation~\cite{Ha2024EOreview}, offering compact footprints~\cite{Reza2016Metalens,Devlin2016TiO2}, enhanced light--matter interaction, and versatile control over the intensity, phase, polarization, and frequency of incident light~\cite{Ahmed2022Review,dorrah2025free}. Mechanisms integrated into metasurfaces for EO modulation include liquid crystals~\cite{Li2019LCKerker,Zou2019LC}, phase-change materials~\cite{Yang2025PhaseChange,Karst2021PhaseChange}, thermo-optic effects~\cite{Sokhoyan2024SiThermoO}, free-carrier plasma dispersion~\cite{Thomaschewski2024ITO,Salary2020PNKerker}, microelectromechanical systems~\cite{Tang2025Moire}, and electro-optic effects~\cite{Benea2022EOPolymer,Soma2025EOPolymer,Chen2025LNO,Dagli2025LNO}. Among these, the linear electro-optic effect, or Pockels effect, originates from the second-order nonlinear susceptibility ($\chi^{(2)}$) of non-centrosymmetric crystals~\cite{Chelladurai2025rMeas} and modulates their refractive indices under an applied quasi-static electric field. Involving no carrier transport, thermal diffusion, or mechanical motion, it offers an inherently ultrafast response, uniquely suited among these mechanisms for high-speed, low-loss EO modulation~\cite{Wang2018LNO}.

Metasurface EO modulators based on the Pockels effect generally rely on nanostructured resonators to confine optical fields within the active material, together with interdigitated electrode (IDE) arrays that deliver strong quasi-static electric fields to the same region~\cite{Benea2022EOPolymer,Soma2025EOPolymer,Dagli2025LNO,Chen2025LNO,DiFrancescantonio2025LNO,Damgaard2023LNO}. Substantial modulation efficiency typically requires closely spaced electrodes (gaps of sub-$\mu$m to a few $\mu$m) to generate a stronger quasi-static field at a given voltage, but the resulting high electrode density increases parasitic resistance and capacitance~\cite{Olthuis1995RC}, limiting the modulation bandwidth to typically sub-GHz. Shrinking the metasurface aperture (e.g., from hundreds of $\mu$m down to tens of $\mu$m) can mitigate these parasitics by reducing both the length and the number of electrodes, yet demands tighter optical focusing, which limits power handling and degrades modulation efficiency through finite-size-induced resonance loss and resonance broadening under finite-angular-spread illumination~\cite{Dolia2024QandSize,Fan2002GMR}. These competing constraints establish fundamental trade-offs between aperture, modulation efficiency, and bandwidth; optimizing all three simultaneously is challenging, unless the Pockels effect is both intrinsically strong and efficiently harnessed, thereby enabling higher efficiency without sacrificing aperture or bandwidth (Table~\ref{tab:comparison}).

Among known EO materials, barium titanate (BaTiO$_3$, or BTO) stands out as a particularly promising candidate for overcoming these trade-offs, owing to its large Pockels coefficient. Its largest Pockels tensor element, $r_{42}=r_{51}$, has been reported to reach $\sim 1300~\mathrm{pm/V}$ in bulk crystals~\cite{Zgonik1994BTO} and $\sim 923~\mathrm{pm/V}$ in thin films grown by molecular beam epitaxy (MBE)~\cite{Abel2019MBEBTO}---far above the largest tensor element of the widely used lithium niobate (LiNbO$_3$, or LNO), $r_{33} \sim 30~\mathrm{pm/V}$~\cite{Weis1985LNO}. Despite this potential, efficiently integrating BTO into metasurfaces remains nontrivial, primarily because thin-film growth and patterning must simultaneously deliver high crystalline quality and well-defined nanostructures. MBE yields high-quality BTO films but is costly and lacks scalability, whereas wet-chemistry routes such as sol--gel processing combined with nanoimprinting are scalable but give limited crystalline quality and hence weaker Pockels effects~\cite{Karvounis2020BTO,Weigand2024BTO,Prountzou2026BTO}. In addition, existing patterning methods such as dry etching are not yet sufficiently mature for fabricating well-defined BTO nanostructures, increasing scattering loss and degrading the resonance quality factor ($Q$ factor)~\cite{Weigand2024BTO,Prountzou2026BTO}. Consequently, no EO metasurface based on vapor-phase deposited BTO has been reported experimentally to date.

Here, we introduce a hybrid BTO/TiO$_2$ metasurface architecture that addresses these limitations. First, high-quality BTO films are obtained via high-throughput radio-frequency (RF) magnetron sputtering; such films have been reported to exhibit effective Pockels coefficients exceeding $100~\mathrm{pm/V}$~\cite{Agham2021BTO,Demkov2024BTO}. Second, rather than directly patterning BTO, we fabricate TiO$_2$ nanostructures on top of a uniform BTO layer, exploiting the more mature and lower-defect nanofabrication of TiO$_2$~\cite{Devlin2016TiO2,Yu2025HighQ,Pernille2026QW}. Leveraging these advantages, we design and experimentally realize resonances with an optical confinement factor of $\sim 0.8$ in the BTO and $Q$ factors exceeding 1300, both beyond the values typically reported for devices based on directly patterned active materials. Third, gold IDEs interleaved with the metasurface efficiently apply direct-current (DC) or RF electric fields to the resonant region in the continuous BTO layer while only weakly perturbing the resonance. The resulting strong overlap between the optical field, the driving electric field, and the BTO enhances the modulation efficiency with little sacrifice in modulation bandwidth or device aperture. A device with a $0.3~\mathrm{mm} \times 0.3~\mathrm{mm}$ metasurface achieves a modulation efficiency of $\sim 0.020~\mathrm{V^{-1}}$ and a $-3~\mathrm{dB}$ bandwidth of $\sim 0.8~\mathrm{GHz}$. Across the tested devices, the effective Pockels coefficients extracted for the BTO span $122$--$151~\mathrm{pm/V}$. Our hybrid BTO/TiO$_2$ architecture thus offers a viable and scalable route for integrating BTO into versatile free-space EO systems.

\section{Results}

\subsection{Design and Fabrication of the Hybrid BTO/TiO$_2$ Metasurface Electro-Optic Modulator}

\begin{figure}
\centering
\includegraphics[width=1\linewidth]{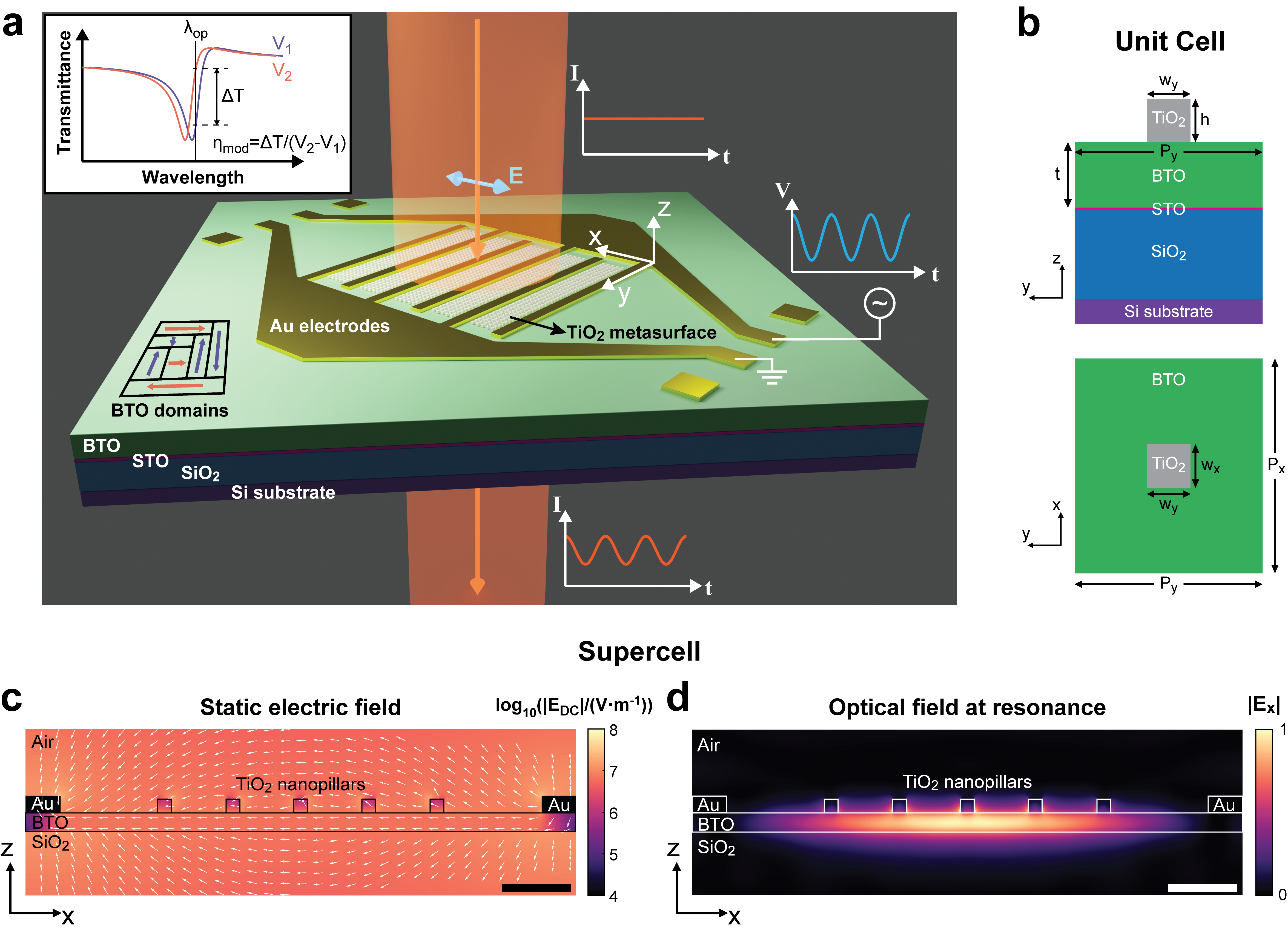}% Here is how to import EPS art
\caption{\textbf{Working principle of the hybrid BTO/TiO$_{2}$ metasurface electro-optic modulator.} \textbf{a}, Schematic of the hybrid BTO/TiO$_2$ metasurface electro-optic modulator. The coordinate axes define the spatial coordinate system used throughout this work. The incident light is $x$-polarized. A time-varying voltage applied across the gold electrodes modulates the intensity of transmitted light. The upper-left inset illustrates the resonance-based modulation mechanism. The inset labeled ``BTO domains'' depicts two types of ferroelectric domains with mutually orthogonal $c$-axis orientations, indicated by the orange and purple arrow colors. Before electrical poling, the spontaneous ferroelectric polarization in each domain type can point along either of two antiparallel directions, represented by oppositely oriented arrows of the same color. \textbf{b}, Side view (top) and top view (bottom) of the unit cell of the metasurface. \textbf{c}, Cross-sectional static electric field profile of a supercell with a 50~V bias in the $x$--$z$ plane through the centers of TiO$_2$ nanopillars. White arrows represent the direction of the field. \textbf{d}, Cross-sectional optical field profile of $E_{\mathrm{x}}$ magnitude (in arbitrary unit) at the resonance wavelength in the same $x$--$z$ plane as in \textbf{c}. The optical field at resonance is well confined within the BTO layer. In \textbf{c} and \textbf{d}, the scale bars are 1 $\mathrm{\mu m}$.}
\label{fig:fig1}
\end{figure}

\begin{figure}
\centering
\includegraphics[width=1\linewidth]{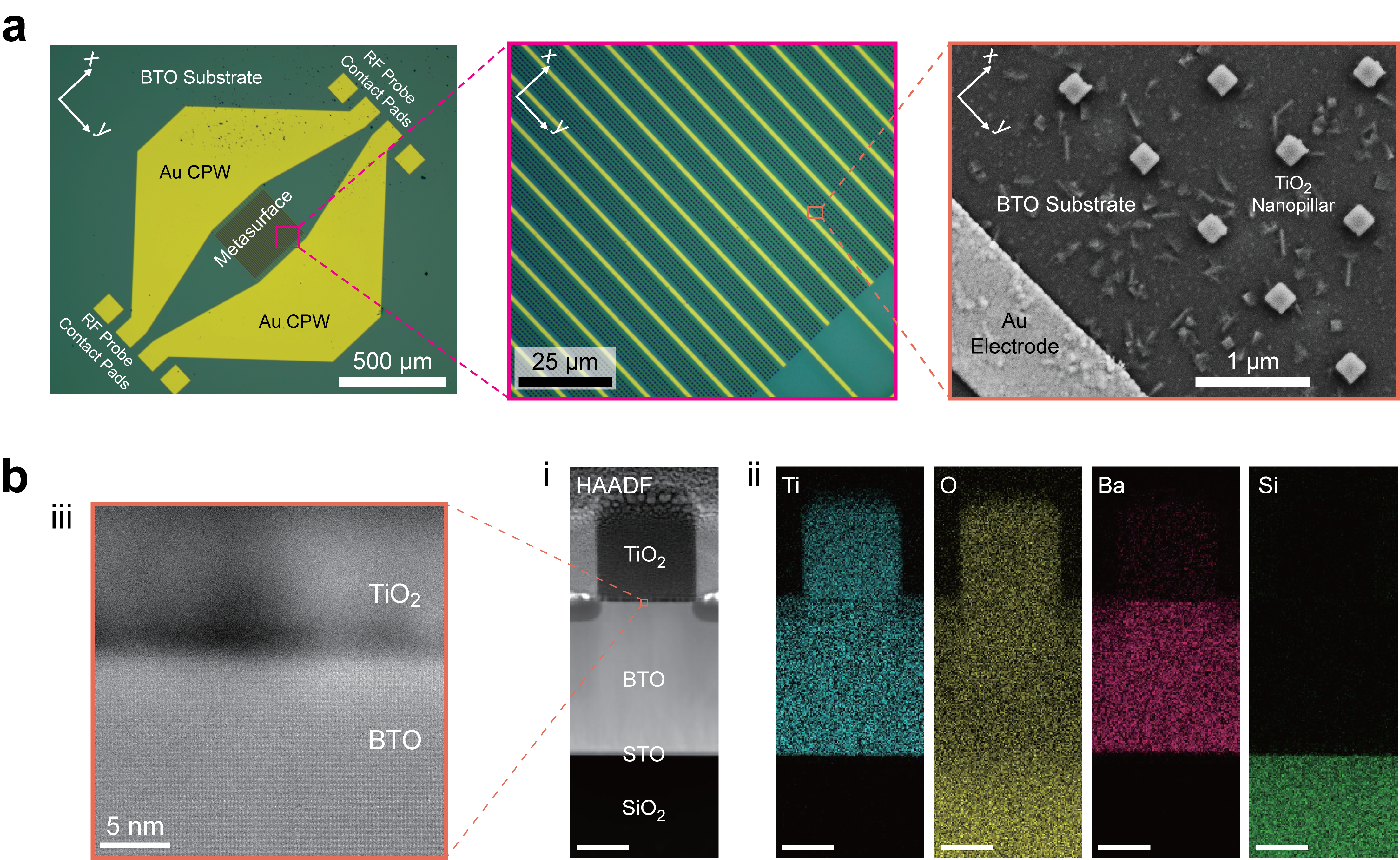}% Here is how to import EPS art
\caption{\textbf{Microscopy images of fabricated devices.} \textbf{a}, Low- (left) and high-magnification (middle) optical images and SEM image (right) of a device with a $0.3~\mathrm{mm} \times 0.3~\mathrm{mm}$ metasurface. The images show well-aligned and well-defined structures. \textbf{b}, Low- (i) and high-magnification (iii) cross-sectional high-angle annular dark-field scanning transmission electron microscopy (HAADF-STEM) images and corresponding elemental maps of Ti, O, Ba, and Si (ii) for a TiO$_2$ nanopillar and the underlying films. Well-defined lattice fringes are observed in the BTO layer in (iii), confirming its retained crystalline quality after fabrication. Scale bars in (i) and (ii), 100~nm.}
\label{fig:fig2}
\end{figure}

The hybrid BTO/TiO$_2$ metasurface modulator consists of a BTO/SrTiO$_3$ (STO)/SiO$_2$/Si multilayer substrate, a TiO$_2$ metasurface, and Au electrodes, as shown in Fig.~1a. The upper-left inset illustrates the operating principle: an optical resonance results in a spectrally narrow transmission dip. When an external electric field varies the refractive index of BTO via the Pockels effect, the resonance condition changes, leading to a spectral shift of the transmission dip. At a fixed operating wavelength $\lambda_{\mathrm{op}}$ on the slope of the resonance, different applied voltages modulate the transmittance. The modulation efficiency $\eta_{\mathrm{mod}}$ of such a device is defined as the absolute change in transmittance per volt under DC conditions at $\lambda_{\mathrm{op}}$, quantifying the transmitted power modulation per unit voltage and input optical power. Achieving a high $\eta_{\mathrm{mod}}$ requires a strong electro-optic interaction, which demands maximal spatial overlap between the BTO layer, the optical field, and the quasi-static electric field.
 
As the basis of the device, we deposit a 280-nm-thick BTO film by RF magnetron sputtering on a 6.4-nm-thick STO buffer layer atop a 3-$\mu$m-thick SiO$_2$ layer on a silicon substrate. At room temperature, BTO is ferroelectric with a tetragonal lattice whose unit cell has two shorter $a$ axes and one longer $c$ axis, also known as the polar axis~\cite{Abel2013BTOPole,Castera2016BTOPole,Abel2019MBEBTO}. An X-ray diffraction scan of our BTO film (Supplementary Fig.~S1) indicates that the $c$ axis lies in the $x$--$y$ plane (see the coordinate frame in Fig.~1a). Because of lattice matching during growth, the in-plane $c$ axes are constrained to two mutually orthogonal orientations, giving four possible ferroelectric polarization directions after growth, all lying at $45^{\circ}$ to $\hat{\mathbf{x}}$ (arrows in the ``BTO domains'' inset of Fig.~1a)~\cite{Dong2023BTO}. Within each domain type, the antiparallel polarizations give opposite Pockels responses that largely cancel the net refractive index change. This issue can be addressed by applying a strong in-plane static electric field to align the antiparallel polarizations into a single direction before measurement---a process known as poling. To maximize the Pockels effect in our BTO, the poling field (and subsequently the optical and driving electric fields) should point along $\pm\hat{\mathbf{x}}$ or $\pm\hat{\mathbf{y}}$ to ensure that both domain types are poled and contribute to the electro-optic modulation~\cite{Abel2019MBEBTO,Dong2023BTO}.

We then design a TiO$_2$ metasurface and Au electrodes on top of the BTO with the orientations shown in Fig.~1a. We first consider the dielectric metasurface alone. The unit cell geometry of the metasurface (Fig.~1b) features cubic pillars ($w_x = w_y = h = 200~\mathrm{nm}$) arranged in a rectangular array ($P_x = 990~\mathrm{nm}$, $P_y = 865~\mathrm{nm}$) on the uniform BTO layer of thickness $t = 280~\mathrm{nm}$. Here we exploit a guided-mode resonance (GMR) to concentrate light within the BTO. The $y$-periodicity of the metasurface array diffracts the incident light into a transverse-electric (TE) leaky mode guided by the effective high-index BTO/TiO$_2$ slab, momentarily trapping it in the plane; as this mode propagates, it continuously re-radiates into free space, and the interference between this leakage and the directly transmitted light produces a sharp, high-$Q$ resonance at a wavelength of $\sim 1580~\mathrm{nm}$ (see Supplementary Note~1 for details). 

We now incorporate the Au electrodes, which apply the driving field to the BTO. They consist of two large coplanar-waveguide (CPW) electrodes, from which IDEs extend into the metasurface region (Fig.~1a). The electrodes are 240~nm thick. The IDEs are 1~$\mu$m wide, with a center-to-center spacing (pitch) of 8~$\mu$m, corresponding to a 7~$\mu$m gap. Within each gap, a metasurface sub-array with five unit cells along $x$ spans $5P_{x}=4.95~\mu$m, leaving a $\sim 1~\mu$m separation from the IDE on each side to mitigate GMR-mode loss from metal absorption.

To investigate the electrostatic and optical responses of the full system, we define a supercell that spans between the centers of two adjacent IDEs along $x$ and retains the $P_{y}$ periodicity along $y$. The simulated static and optical field profiles of the supercell in an $x$--$z$ cross-section through the centers of the TiO$_2$ nanopillars are shown in Fig.~1c,d, respectively. To quantitatively evaluate the field overlap with the BTO layer, we define a three-dimensional active region as the BTO contained within the five metasurface unit cells. In this region, the static electric field is aligned along the $x$ direction (Fig.~1c; the spatially averaged squared magnitudes of $E_y$ and $E_z$ are $\sim 10^{-6}$ of that of $E_x$) and varies in magnitude by less than $0.5\%$. Accordingly, we approximate the static field in the active region as a uniform field along $x$. Under a 50~V bias, the average static field $E_{\mathrm{DC}}$ in the active region is $6.92~\mathrm{V/\mu m}$, close to the ideal maximum of $7.14~\mathrm{V/\mu m}$ (50~V divided by the 7~$\mu$m gap). This efficient use of the bias arises because the continuous BTO layer, whose permittivity is on the order of $10^{3}$~\cite{Zgonik1994BTO,Chelladurai2025rMeas}, bridges the two IDEs without interruption. In patterned-BTO devices, by contrast, any low-permittivity material present in series can drop most of the applied voltage, resulting in inefficient use of the bias~\cite{Prountzou2026BTO,Weigand2024BTO}.
 
The previously identified TE GMR is also observed in the supercell simulation. The optical electric field at resonance (Fig.~1d) is predominantly polarized along the $x$ direction and well confined within the active region, with a large confinement factor $\Gamma_{\mathrm{AR}} \approx 0.8$ (see Methods). This large confinement factor stems in part from the unpatterned BTO layer, which sustains continuous confinement unattainable with a patterned BTO. In addition, because BTO and TiO$_2$ are closely index-matched, the field is not pinned in the nanopillars (as it would be with a high-contrast material such as silicon~\cite{Dagli2025LNO}), so the mode profile can instead be shaped by design. Together, these simulations demonstrate that our hybrid BTO/TiO$_2$ architecture enables strong spatial overlap between the optical and driving fields within the BTO, ideal for enhancing the modulation efficiency $\eta_{\mathrm{mod}}$.

The fabricated devices show excellent agreement with the design, with precise alignment between the Au structures and TiO$_2$ nanopillars (Fig.~2a). A key feature of our fabrication process is that the BTO layer remains fully covered by the resist and TiO$_2$ during the TiO$_2$ dry etch (Supplementary Note~2), such that its surface stays intact and its high crystalline quality is preserved---as confirmed by the smooth interfaces, sharp elemental contrast, and clear BTO lattice fringes in Fig.~2b. This device quality, enabled by our choice of materials and fabrication approach, allows the designed performance to be realized experimentally.

\subsection{DC Electro-Optic Characterization of the Hybrid BTO/TiO$_2$ Metasurface Modulator}

\begin{figure}
\centering
\includegraphics[width=1\linewidth]{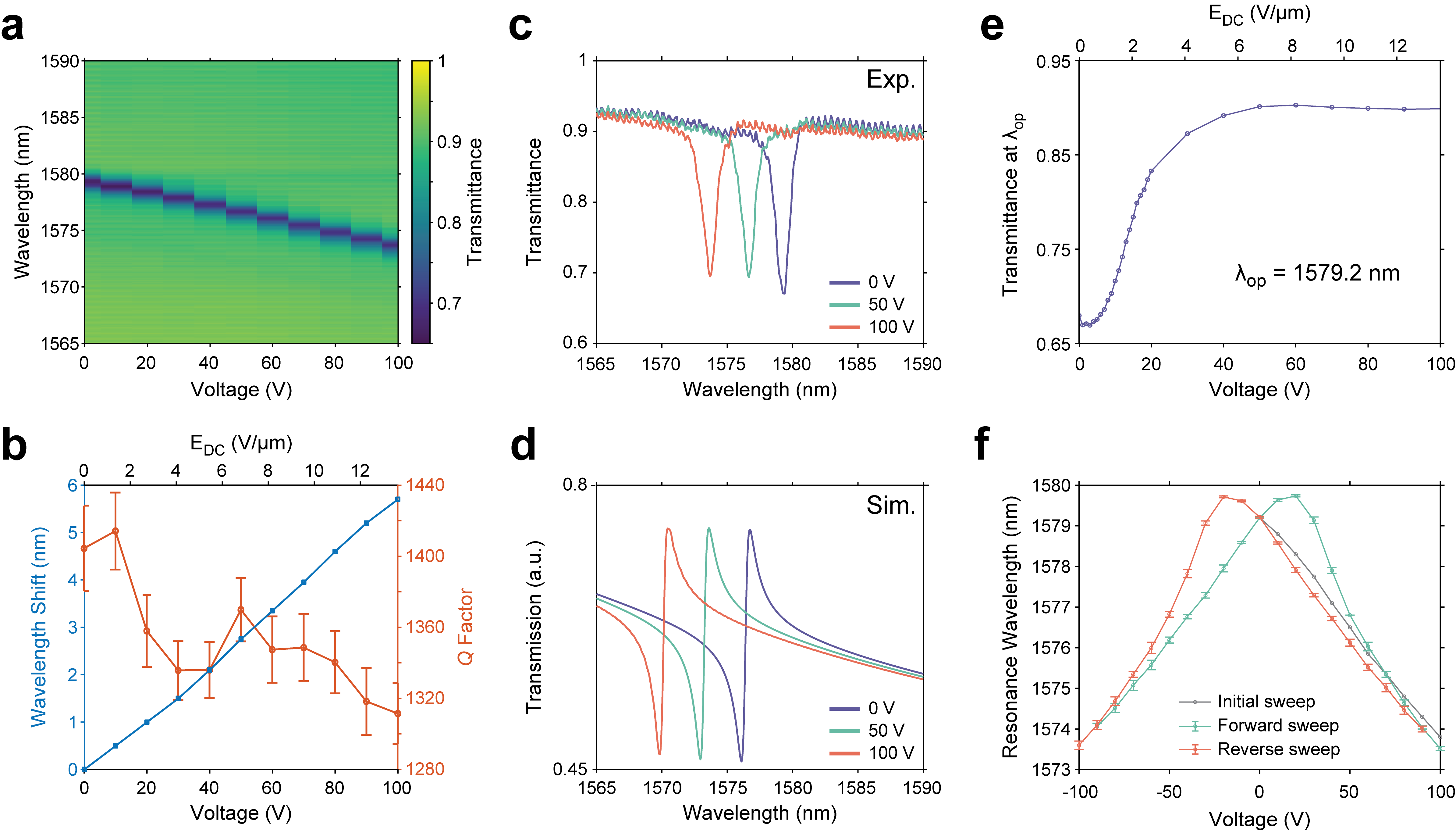}% Here is how to import EPS art
\caption{\textbf{DC characterization of a device with a $2~\mathrm{mm} \times 2~\mathrm{mm}$ metasurface.} \textbf{a}, Measured transmittance map under a DC bias swept from 0 to 100~V. \textbf{b}, Resonance wavelength shift extracted from the spectra in \textbf{a} and $Q$ factor extracted from Fano fits to the spectra. Error bars represent the standard errors from the Fano fits. \textbf{c}, Transmittance spectra selected from \textbf{a} at 0, 50, and 100~V. \textbf{d}, Corresponding simulated transmission spectra at 0, 50, and 100~V from the supercell simulations, without substrate normalization. The effective Pockels coefficient of BTO, $r_{\mathrm{eff,sim}}$, is assumed to be $130~\mathrm{pm/V}$ in the simulations. \textbf{e}, Measured transmittance at the operating wavelength ($\lambda_{\mathrm{op}}$) as a function of voltage. Here, $\lambda_{\mathrm{op}}$ is chosen as the zero-bias resonance-dip wavelength (1579.2~nm). Transmittance in \textbf{a}, \textbf{c}, and \textbf{e} is normalized to the spectrum of the bare substrate. Top $x$ axes in \textbf{b} and \textbf{e} show the static electric field in the BTO active region extracted from the electrostatic simulations. \textbf{f}, Measured resonance-dip wavelength as a function of the applied voltage during eight repeated voltage cycles, revealing a ferroelectric hysteresis loop. In each cycle, the voltage was swept from 0 to $+100$~V, then to $-100$~V, and finally back to 0~V. The gray line denotes the initial sweep from 0 to $+100$~V, whereas the orange and green lines show the mean responses during the subsequent reverse and forward sweeps, respectively. Error bars represent the standard deviations across repeated sweeps.} 
\label{fig:fig3}
\end{figure}

We experimentally characterize the static modulation performance of a device with a large ($2~\mathrm{mm} \times 2~\mathrm{mm}$) metasurface under normally incident, $x$-polarized illumination. Prior to measurement, the device was poled by applying a 50~V bias for 15~min. The voltage was subsequently swept from 0 to 100~V, and the corresponding transmission spectra were recorded. As shown in Fig.~3a,b, the resonance appears within the targeted wavelength range, exhibiting a transmittance contrast of $\sim 0.25$, and Fano fits yield $Q$ factors of $\sim 1370$. With increasing voltage, the resonance blueshifts continuously. Fig.~3c,d compares experimental spectra at 0, 50, and 100~V with the results from the supercell simulations. In the simulations, we incorporate the Pockels effect as a perturbation to the BTO refractive index $n_{\mathrm{BTO}}$ governed by~\cite{BoydNonlinear}
\begin{equation}
\Delta\left(\frac{1}{n_{\mathrm{BTO}}^2}\right) = r_{\mathrm{eff,sim}} \, E_{\mathrm{DC}},
\label{eq:Pockels} 
\end{equation}
where $r_{\mathrm{eff,sim}} = 130~\mathrm{pm/V}$ is the assumed effective Pockels coefficient and $E_{\mathrm{DC}}$ is the applied static electric field obtained from the electrostatic simulations ($6.92~\mathrm{V/\mu m}$ for 50~V and $13.84~\mathrm{V/\mu m}$ for 100~V). The simulated spectra show a blueshift similar to that observed experimentally, but with a more pronounced Fano asymmetry, as is typical for GMRs~\cite{Fan2002GMR}. We attribute the weaker asymmetry in the measurements to residual loss from device imperfections (Supplementary Note~5).

From Fig.~3b, we extract the resonance wavelength shift rate $S_{\mathrm{exp}} = \left| \Delta \lambda_{\mathrm{res,exp}} / \Delta V \right| \approx 0.057~\mathrm{nm/V}$. Using this value, the experimental effective Pockels coefficient $r_{\mathrm{eff,exp}}$ is estimated by comparison with simulations and by using perturbation theory~\cite{Benea2022EOPolymer,Soma2025EOPolymer,PhCBook} (see Methods for details), yielding $r_{\mathrm{eff,exp}} \approx 118~\mathrm{pm/V}$ and $126~\mathrm{pm/V}$, respectively. The two
independently obtained estimates are in good agreement. The effective Pockels coefficient is approximately four times the intrinsic $r_{33}$ of LiNbO$_3$. This large effective Pockels coefficient reflects the high quality of both the BTO film and the fabricated device, as well as the effectiveness of the poling process.

The viability of the device for practical applications is quantified by the device-level modulation efficiency $\eta_{\mathrm{mod}}$. With $\lambda_{\mathrm{op}}$ set to the zero-bias resonance-dip wavelength (1579.2~nm), a maximum $\eta_{\mathrm{mod}} \approx 0.014~\mathrm{V^{-1}}$ is achieved over the voltage range of 9--16~V (Fig.~3e). This $\eta_{\mathrm{mod}}$ is among the highest reported for metasurface EO modulators (Table~\ref{tab:comparison}). This $\sim 7~\mathrm{V}$ operating range can be centered around 0~V by choosing $\lambda_{\mathrm{op}}$ at the point of maximum slope of the resonance. Notably, the electrode gap in this work is relatively large compared with previous studies. Reducing the electrode spacing can further improve $\eta_{\mathrm{mod}}$, thereby enabling full-swing operation at voltage levels compatible with complementary metal--oxide--semiconductor (CMOS) electronics.

Finally, we investigate the ferroelectric hysteresis behavior of our BTO. The device was first poled at $50~\mathrm{V}$ for 15~min, followed by a voltage sweep from $0$ to $100~\mathrm{V}$, then to $-100~\mathrm{V}$, and back to $0$. This cycle was repeated eight times, and the spectral response was recorded at each step. As shown in Fig.~3f, the resonance-dip wavelength exhibits a characteristic butterfly hysteresis loop. The wavelength maxima occur near the polarization-switching midpoint, providing an optical estimate of the coercive field. The extracted coercive voltage is $\sim 20~\mathrm{V}$, corresponding to a coercive field of $\sim 2.8~\mathrm{V/\mu m}$, consistent with previously reported values for thin-film BTO~\cite{Abel2013BTOPole,Abel2019MBEBTO,Dong2023BTO}. The relatively low coercive field suggests that applying an additional DC bias during modulation can help stabilize the electro-optic response by suppressing spontaneous polarization reversal.

\subsection{GHz-Speed Electro-Optic Modulation with the Hybrid BTO/TiO$_2$ Metasurface}

\begin{figure}
\centering
\includegraphics[width=1\linewidth]{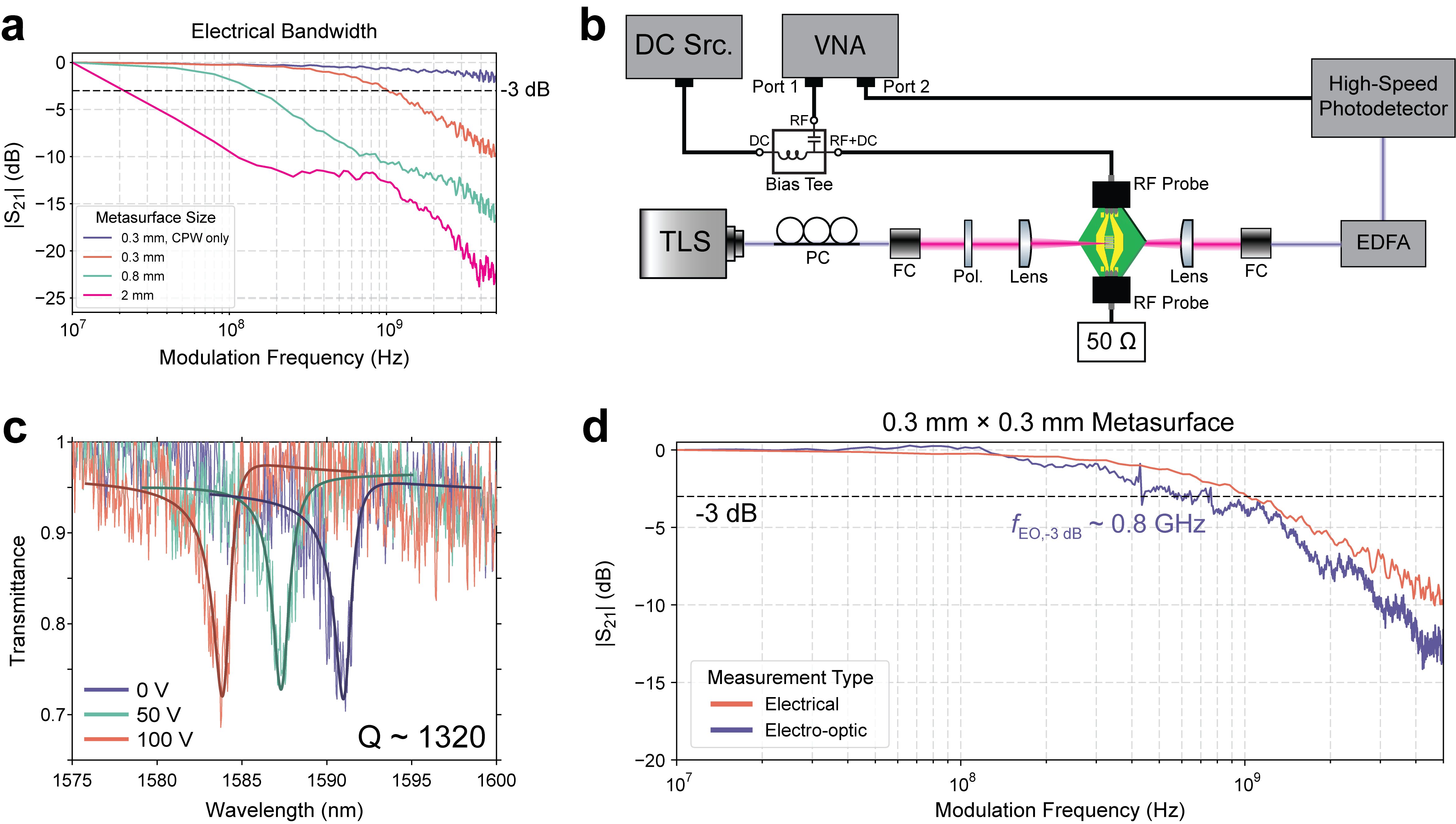}% Here is how to import EPS art
\caption{\textbf{GHz-speed modulation characterization.} \textbf{a}, High-speed electrical modulation responses of devices with different metasurface lateral sizes. The purple line corresponds to a CPW-only reference without IDEs or metasurfaces. The electrode thickness in all these devices is $55~\mathrm{nm}$, rather than the $240~\mathrm{nm}$ used before. The $S_{21}$ response was measured using a vector network analyzer. RF cable losses were de-embedded. \textbf{b}, Measurement setup for high-speed electro-optic modulation. A $50~\Omega$ resistor is connected to the bottom RF probe, forming a terminated RF circuit. VNA, vector network analyzer; DC Src., DC voltage source; TLS, tunable laser source; PC, polarization controller; FC, fiber collimator; Pol., polarizer; EDFA, erbium-doped fiber amplifier. \textbf{c}, Measured transmittance spectra of the $0.3~\mathrm{mm} \times 0.3~\mathrm{mm}$ device in \textbf{a} at 0, 50, and 100~V DC biases. Spectra are normalized to that of the bare substrate. Light thin lines show the experimental data, while dark thick lines show the corresponding Fano fits. The fitted $Q$ factors are $\sim 1320$. \textbf{d}, High-speed modulation responses of the same device as in \textbf{c}. The $-3~\mathrm{dB}$ electro-optic bandwidth $f_{\mathrm{EO},-3\,\mathrm{dB}}$ reaches $\sim 0.8~\mathrm{GHz}$, demonstrating GHz-speed electro-optic modulation. RF cable losses were de-embedded.}
\label{fig:fig4}
\end{figure}

As discussed previously, the modulation bandwidth of devices with IDEs is typically RC-limited~\cite{Benea2022EOPolymer,Soma2025EOPolymer,Dagli2025LNO}. To determine the largest metasurface aperture compatible with GHz-speed modulation, we use a vector network analyzer (VNA) to measure the electrical $S_{21}$ responses of devices with different metasurface sizes, together with a CPW-only reference without IDEs or metasurfaces (Fig.~4a). The latter exhibits the flattest $\left|S_{21}\right|$ response across the measured RF frequency range. As the metasurface size increases, the $-3~\mathrm{dB}$ electrical bandwidth decreases, confirming the expected RC limitation. These results suggest that GHz-speed EO modulation of our devices is feasible with metasurface sizes up to approximately $0.3~\mathrm{mm} \times 0.3~\mathrm{mm}$. We note that, for these devices, the electrical bandwidth can be further improved by reducing the device capacitance, whereas the electrode resistance remains sub-dominant even at the reduced electrode thickness used here ($55~\mathrm{nm}$; Supplementary Note~7).

To characterize the electro-optic modulation speed of the $0.3~\mathrm{mm} \times 0.3~\mathrm{mm}$ device, we employ the setup shown in Fig.~4b. We first measure its voltage-dependent transmission under DC bias, with all RF components removed (Fig.~4c and Supplementary Fig.~S10). Consistent with the $2~\mathrm{mm} \times 2~\mathrm{mm}$ device in Fig.~3, the resonance exhibits a blueshift. Fits to the Fano lineshape yield quality factors of $Q \sim 1320$ and a transmittance contrast of $\sim 0.24$ under 0~V bias, both comparable to the corresponding values extracted from the $2~\mathrm{mm} \times 2~\mathrm{mm}$ device. The extracted resonance shift rate, $S_{\mathrm{exp}} \approx 0.071~\mathrm{nm/V}$, and the maximum modulation efficiency, $\eta_{\mathrm{mod}} \approx 0.020~\mathrm{V^{-1}}$, are higher than those of the $2~\mathrm{mm} \times 2~\mathrm{mm}$ device. The difference in electrode thickness between the two devices has a negligible effect on the static and optical fields in the active region (Supplementary Note~3), so the simulated $E_{\mathrm{DC}}$ and $\Gamma_{\mathrm{AR}}$ remain applicable. This larger shift rate then corresponds to a larger effective Pockels coefficient of $r_{\mathrm{eff}} \approx 151~\mathrm{pm/V}$. The enhanced $r_{\mathrm{eff}}$ value may be associated with device-to-device variations in BTO crystalline quality (the two devices were from different wafers), device orientation, or poling conditions.

Based on the DC measurements above, we set the operating DC bias to $7~\mathrm{V}$ to maintain a stable ferroelectric polarization state and set the operating wavelength to the point of maximum slope of the resonance for dynamic modulation measurements. A bias tee combines the DC bias with the RF signal from the VNA, and a $50~\Omega$ resistor provides circuit termination. The measured electro-optic response for the $0.3~\mathrm{mm} \times 0.3~\mathrm{mm}$ device is shown in Fig.~4d. GHz-speed modulation is observed, with a $-3~\mathrm{dB}$ electro-optic bandwidth, $f_{\mathrm{EO},-3\,\mathrm{dB}}$, of $\sim 0.8~\mathrm{GHz}$. The EO bandwidth is only slightly smaller than the electrical bandwidth ($\sim 1~\mathrm{GHz}$). This difference may arise from RF-frequency dispersion of $r_{\mathrm{eff}}$, including the dispersion inherited from $\chi^{(1)}$ through Miller's rule~\cite{Chelladurai2025rMeas} and a reduced acoustic/strain-mediated contribution at high frequencies~\cite{kim2023nature}. We note that the $7~\mathrm{V}$ DC bias across the $50~\Omega$ termination dissipates $\sim 1~\mathrm{W}$ of static power, which AC-coupling would eliminate (Supplementary Note~7).

\section{Discussion}

\begin{table}[htbp]
\caption{Performance comparison of representative metasurface electro-optic modulators}
\label{tab:comparison}
\centering
\small
\setlength{\tabcolsep}{4pt}
\begin{tabular}{lcccccc}
\toprule
References &
EO material &
\makecell[c]{Metasurface size\\($\mu\mathrm{m}^2$)} &
\makecell[c]{DC $\eta_{\mathrm{mod}}$\tnote{a}\\($\mathrm{V}^{-1}$)} &
\makecell[c]{$f_{\mathrm{EO},-3\,\mathrm{dB}}$\tnote{b}\\(GHz)} &
\makecell[c]{$Q$ factor\tnote{c}} &
\makecell[c]{$r_{\mathrm{eff}}$\tnote{d}\\(pm/V)} \\
\midrule
This work & BTO & $300\times300$ & 0.020 & 0.8 & 1320 & 151 \\
Ref.~\cite{Prountzou2026BTO} & BTO & $50\times50$ & 0.00053 & 0.001 & 200 & --- \\
Ref.~\cite{Chen2025LNO} & LNO & $200\times200$ & 0.0023 & $5\times10^{-7}$ & 440 & --- \\
Ref.~\cite{Dagli2025LNO} & LNO & $160\times160$ & 0.0071 & 0.34 & 2200 & --- \\
Ref.~\cite{Soma2025EOPolymer} & EO polymer & $10\times10$ & 0.061 & 0.38 & 655 & 58 \\
Ref.~\cite{Fukui2025EOPolymer} & EO polymer & $40\times40$ & $6.2\times10^{-5}$ & 17.5 & 102 & 20 \\
Ref.~\cite{Benea2022EOPolymer} & EO polymer & $330\times330$ & 0.007 & 3 & 550 & --- \\
\bottomrule
\end{tabular}
\begin{tablenotes}
\footnotesize
\item[a] DC modulation efficiency, as defined in the main text.
\item[b] $-3$~dB electro-optic modulation bandwidth.
\item[c] Quality factor of the optical resonance.
\item[d] Effective Pockels coefficient.
\item[] ---, not reported.
\end{tablenotes}
\end{table}

By incorporating RF-magnetron-sputtered BTO thin films into the hybrid BTO/TiO$_2$ architecture that efficiently harnesses their Pockels effect, we achieved, in a single device, a high modulation efficiency ($\eta_{\mathrm{mod}} \sim 0.020~\mathrm{V^{-1}}$), a high modulation bandwidth ($f_{\mathrm{EO},-3\,\mathrm{dB}}\sim 0.8~\mathrm{GHz}$), and a large metasurface aperture ($0.3~\mathrm{mm} \times 0.3~\mathrm{mm}$), thereby overcoming the aforementioned three-way trade-off and placing our device among the best metasurface EO modulators reported to date (Table~\ref{tab:comparison}). Compared with devices based on LNO and solution-derived BTO, our device excels in all three performance dimensions, owing to the much larger effective Pockels coefficient and outstanding resonance properties---the high $Q$ factor and large confinement factor $\Gamma_{\mathrm{AR}}$. Although EO polymers have intrinsic Pockels coefficients comparable to those of BTO~\cite{Benea2021EOP,Jin2014EOP}, their effective values in devices are often substantially lower, and the materials can suffer from instability and degradation~\cite{Feng2024EOPInsta}. With careful design, such devices can reach record performance in a single dimension at the expense of the others. In contrast, we improved all three dimensions simultaneously for broad applicability, while our platform retains the potential to push any single dimension to a record value for specific applications.

The results also point to opportunities for further improvement. At present, the modulation efficiency is mainly limited by the transmittance contrast, which is reduced by the presence of the metallic electrodes (Supplementary Note~1). One route to improvement is to co-optimize the metasurface geometry and electrode layout, maximizing the lateral extent of the resonant region while minimizing its modal overlap with the lossy metallic electrodes. A complementary route is to replace the metal with low-loss conductive electrodes such as heavily doped n-InP or n-Si, which combine high electrical conductivity with low optical absorption at telecommunication wavelengths and could even double as resonant metasurface elements~\cite{Fukui2025EOPolymer}. The modulation speed can likewise be improved by reducing the device capacitance, for example by slightly shrinking the device size or adjusting the BTO thickness and electrode layout, and by improving the impedance matching of the electrodes, potentially pushing the bandwidth beyond $10~\mathrm{GHz}$.

The high-performance BTO-based EO modulators demonstrated in this work hold strong potential for high-speed free-space optical communication, which can be more advantageous than fiber-based links in certain scenarios---most notably inter-satellite communication. Looking forward, our hybrid BTO/TiO$_2$ architecture can be extended to pixelated and phase-modulation devices~\cite{Benea2021EOP,Panuski2022SLM,Park2021Lidar}, enabling spatial light modulation at the GHz level. This would overcome the long-standing speed bottleneck of modulation systems that rely on mechanical motion or liquid crystals, and could transform technologies such as LiDAR and neutral-atom quantum computing.

% =====================
\section{Methods}
% =====================
\subsection{Electrostatic simulation}

Three-dimensional electrostatic simulations were performed on the supercell (Fig.~1c) using the Electrostatics Module in COMSOL Multiphysics (v6.2, COMSOL AB, Stockholm, Sweden), which solves Poisson's equation using the finite element method (FEM). The relative permittivities of silicon, SiO$_2$, and TiO$_2$ used in the simulations are 11.7, 4.2, and 40, respectively. For BTO, a diagonal permittivity tensor is assumed, $\boldsymbol{\epsilon} = \mathrm{diag}(\epsilon_x, \epsilon_y, \epsilon_z)$. In the local crystallographic frame, the permittivity is taken to be $\boldsymbol{\epsilon}_\mathrm{local} = \mathrm{diag}(\epsilon_a, \epsilon_a, \epsilon_c) = \mathrm{diag}(1976, 1976, 112)$, obtained from the COMSOL Piezoelectric Materials Library. Because the BTO $c$-axis lies within the $x$--$y$ plane along either of two orthogonal in-plane directions, each in-plane direction averages the $a$- and $c$-axis responses; we therefore approximate the in-plane permittivity as isotropic, $\epsilon_x = \epsilon_y = (\epsilon_a + \epsilon_c)/2 = 1044$, while the out-of-plane direction always corresponds to the $a$-axis, $\epsilon_z = \epsilon_a = 1976$. The simulation domain consists of two supercells arranged along the $x$ direction. A potential of 0~V is applied to the central electrode located at the interface between the two supercells, while a potential of 50~V is applied to the two outer half-electrodes at the edges of the domain. Periodic boundary conditions are imposed along the $x$ and $y$ directions, and zero-charge (Neumann) boundary conditions are applied along the $z$ direction. Convergence tests were performed for the mesh size of the hybrid metasurface region, as well as for the thicknesses of the air (top) and silicon (bottom) layers. The electric field in the active region of a supercell is quantified by extracting the spatially averaged values and corresponding standard deviations using the global evaluation feature in COMSOL. The resulting electric field is used to evaluate the electro-optic modulation in the optical simulations and to assist in the extraction of the effective Pockels coefficients from the experiments.

\subsection{Optical simulations}

Simulations of the unit cell structure (Supplementary Note~1) were performed using the Stanford Stratified Structure Solver (S$^{4}$), a frequency-domain electromagnetic solver based on rigorous coupled-wave analysis (RCWA)~\cite{Liu2012S4}. The silicon layer was omitted in the RCWA simulations, as the influence of the Si/SiO$_2$ interface on the GMR modes is negligible for the purpose of resonance identification. Instead, the SiO$_2$ layer was treated as a semi-infinite substrate. The refractive indices of BTO, SiO$_2$, and TiO$_2$ were obtained from ellipsometry measurements of the samples and used as input for the simulations. Convergence with respect to the number of Fourier harmonics was carefully verified, and 145 harmonics were retained in the final calculations.

Simulations of the supercell structure (Fig.~1d and Fig.~3d) were performed using Tidy3D (Flexcompute Inc.), a cloud-based finite-difference time-domain (FDTD) solver. The refractive indices of BTO, SiO$_2$, and TiO$_2$ were obtained from ellipsometry measurements of the samples and fitted using the pole--residue model for dispersive material modeling. The refractive indices of Si and Au were taken from the built-in material libraries (``Palik\_Lossless''~\cite{Palik1985} for Si and ``JohnsonChristy1972''~\cite{JohnsonChristy1972} for Au). The Pockels effect of BTO was modeled by applying the perturbation defined in Eq.~\eqref{eq:Pockels} to the BTO refractive index data, followed by refitting using the pole--residue model with fitting errors on the order of $10^{-8}$. This scalar perturbation approximation, rather than a full tensorial modulation of the BTO refractive index, is expected to accurately capture the resonance shift behavior because the targeted resonance mode is an $x$-polarized TE mode, whose optical response is predominantly determined by only the $x$-direction projection of the refractive index ellipsoid. Periodic boundary conditions were imposed along the $x$ and $y$ directions, while perfectly matched layers (PMLs) were applied to the $z$ boundaries. Following convergence tests, a mesh resolution exceeding 60 steps per wavelength and a local mesh override with a minimum mesh size of 4~nm in the hybrid metasurface region were adopted.

\subsection{Optical confinement factor of the active region}

The spatial overlap between the resonant optical field and the BTO active region is quantified by the confinement factor~\cite{Benea2021EOP,Soma2025EOPolymer,PhCBook}
\begin{equation}
\Gamma_{\mathrm{AR}} =
\frac{{\iiint_{AR}} \varepsilon_{\mathrm{BTO}} |E_x(\mathbf{r})|^2 \, \mathrm{d}x\mathrm{d}y\mathrm{d}z}
{\iiint_{SC} \varepsilon(\mathbf{r}) |\mathbf{E}(\mathbf{r})|^2 \, \mathrm{d}x\mathrm{d}y\mathrm{d}z},
\label{eq:Gamma}
\end{equation}
where $\mathbf{E}(\mathbf{r})$ is the electric field of the resonant optical mode at position $\mathbf{r}$ and $E_x(\mathbf{r})$ is its component along $x$; the numerator is integrated over the active region (AR), which is entirely filled with BTO of permittivity $\varepsilon_{\mathrm{BTO}}$, and the denominator over the whole supercell (SC), whose permittivity $\varepsilon(\mathbf{r})$ varies spatially. This ratio gives the fraction of the total electric field energy stored in the supercell that resides in the $x$-polarized field component within the active region. The field distributions were taken from the converged supercell FDTD simulation at the resonance wavelength, and both integrals were evaluated numerically on the simulation grid, yielding $\Gamma_{\mathrm{AR}} \approx 0.8$ for our design.

\subsection{BTO film growth}

Standard photonic-grade silicon-on-insulator (SOI) wafers from Soitec were used as the substrates for BTO thin-film growth. Prior to BTO deposition, a 6.4-nm-thick STO buffer layer was deposited on the cleaned Si surface using molecular beam epitaxy (MBE). The top Si layer of the SOI substrate was then oxidized through the STO layer, thereby merging it with the buried oxide layer. The BTO film was subsequently deposited by off-axis RF magnetron sputtering at a substrate temperature of 700~$^\circ$C to a nominal thickness of 300~nm. Spectroscopic ellipsometry on the final film yields 280~nm, which is the value used in all simulations. The crystallographic orientation and crystalline quality of the BTO film were confirmed using in situ reflection high-energy electron diffraction (RHEED) and ex situ X-ray diffraction (XRD).

\subsection{Device fabrication}

The device was fabricated on top of the BTO film using a two-step process: (i) fabrication of the Au structures, followed by (ii) fabrication of the TiO$_{2}$ nanopillars aligned with the Au structures. The fabrication protocol is detailed in Supplementary Note~2.

\subsection{Scanning transmission electron microscopy}

TEM lamellae were prepared using a focused ion beam (FIB, TESCAN GAIA3) operated at 30~kV for coarse milling, followed by final thinning at 2--5~kV to minimize surface damage and amorphization. HAADF-STEM and EDS were performed using a JEOL JEM-ARM300F transmission electron microscope at 300~kV. The probe convergence semi-angle was set to 25.7~mrad, and the inner collection angle for HAADF imaging was 70~mrad.

\subsection{DC electro-optic characterization of the devices}

Optical transmission spectra were measured at normal incidence using a tunable laser source (TSL-550, Santec) and an optical power meter (MPM-210, Santec). Voltage bias was supplied using a source measure unit (Keithley 2400 SourceMeter). The incident polarization was controlled using a fiber-based polarization controller and a free-space polarizer to selectively excite the $x$-polarized TE resonance mode. Optical lenses were used only when measuring the smaller $0.3~\mathrm{mm} \times 0.3~\mathrm{mm}$ device. Before each measurement session, the device was poled by applying a 50~V bias for $\sim15$~min. For each sample, the measured transmission spectra were normalized to that measured from a bare substrate region on the same sample. Note that the $0.3~\mathrm{mm} \times 0.3~\mathrm{mm}$ device (Fig.~4c) shows stronger spectral fluctuations than the $2~\mathrm{mm} \times 2~\mathrm{mm}$ device (Fig.~3c) in measurements, which may arise from reduced spatial averaging of Fabry--Perot fringes in the silicon layer over the smaller device area. 

\subsection{Spectrum fitting and parameter extraction}

The resonance quality factors were extracted from Fano fits to the measured transmittance spectra. For the $2~\mathrm{mm} \times 2~\mathrm{mm}$ device, the modulation efficiency was directly extracted from the voltage-dependent transmittance at a fixed operating wavelength. For the $0.3~\mathrm{mm} \times 0.3~\mathrm{mm}$ device, the modulation efficiency was calculated from the Fano-fitted resonance slope and the measured resonance shift rate.

The detailed fitting procedure and extraction methods are described in Supplementary Note~6.

\subsection{Extraction of the experimental effective Pockels coefficient $r_{\mathrm{eff,exp}}$ of BTO}

We used two independent approaches to estimate the experimental effective Pockels coefficient $r_{\mathrm{eff,exp}}$ of our BTO. The following example analysis is based on the results in Fig.~3.

First, we estimated $r_{\mathrm{eff,exp}}$ by comparing the experimentally measured resonance shift rate with the simulated value. The difference between $r_{\mathrm{eff,sim}}$ and $r_{\mathrm{eff,exp}}$ leads to a corresponding difference in the refractive index perturbation of BTO, given by 
\begin{equation}
\Delta n_{\mathrm{BTO}} = -\tfrac{1}{2} n_{\mathrm{BTO}}^{3} r_{\mathrm{eff}} E_{\mathrm{DC}}, 
\label{eq:Delta_n}
\end{equation}
as derived from Eq.~\eqref{eq:Pockels}. Under first-order electromagnetic perturbation theory~\cite{PhCBook}, the induced perturbation in the effective refractive index of the guided mode satisfies
\begin{equation}
\Delta n_{\mathrm{eff}}=\Gamma_{\mathrm{AR}}\Delta n_{\mathrm{BTO}}.
\label{eq:delta_neff}
\end{equation}
For guided-mode resonances under normal incidence, the bare resonance wavelength $\lambda_{\mathrm{res}}$, neglecting coupling between counter-propagating modes, can be estimated from the phase-matching condition~\cite{641320}
\begin{equation}
n_{\mathrm{eff}}=\left| m \frac{\lambda_{\mathrm{res}}}{\Lambda} \right|,
\label{eq:GMR}
\end{equation}
where $m$ is the grating diffraction order that excites the target resonance, and $\Lambda$ is the grating period in the diffraction direction. Taking material and modal dispersion into account, first-order perturbation analysis of Eq.~\eqref{eq:GMR} yields
\begin{equation}
\frac{\Delta \lambda_{\mathrm{res}}}{\lambda_{\mathrm{res}}}
=
\frac{\Delta n_{\mathrm{eff}}}{n_g},
\label{eq:GMRdiff}
\end{equation}
where $n_g$ is the group refractive index of the guided mode. The perturbation terms in Eqs.~\eqref{eq:Delta_n}, \eqref{eq:delta_neff}, and \eqref{eq:GMRdiff} are linearly related. Therefore, if the small difference in the resonance wavelength shift rates $S$ between simulations and experiments is assumed to be dominated by a small difference between $r_{\mathrm{eff,sim}}$ and $r_{\mathrm{eff,exp}}$, we have that $S$ scales linearly with $r_{\mathrm{eff}}$, i.e.,
\begin{equation}
\frac{S_{\mathrm{exp}}}{r_{\mathrm{eff,exp}}} = \frac{S_{\mathrm{sim}}}{r_{\mathrm{eff,sim}}}.
\label{eq:SimExp}
\end{equation}
From Fig.~3b--d, we obtained $S_{\mathrm{exp}} \approx 0.057~\mathrm{nm/V}$ and $S_{\mathrm{sim}} \approx 0.063~\mathrm{nm/V}$. With $r_{\mathrm{eff,sim}}=130~\mathrm{pm/V}$, we derived $r_{\mathrm{eff,exp}} \approx 118~\mathrm{pm/V}$.

Another independent estimate was obtained based on perturbation theory, which gives the approximate resonance wavelength shift relation~\cite{Benea2022EOPolymer,Soma2025EOPolymer,PhCBook}
\begin{equation}
\frac{\Delta \lambda_{\mathrm{res}}}{\lambda_{\mathrm{res}}}
\approx
\Gamma_{\mathrm{AR}} \frac{\Delta n_{\mathrm{BTO}}}{n_{\mathrm{BTO}}}.
\label{eq:Shift}
\end{equation}
 Substituting Eq.~\eqref{eq:Delta_n}, $\Gamma_{\mathrm{AR}} = 0.8$, $n_{\mathrm{BTO}} = 2.2785$, $\lambda_{\mathrm{res}}=1579.2$~nm, $\Delta \lambda_{\mathrm{res}}=5.7$~nm, and $E_{\mathrm{DC}}=13.84~\mathrm{V/\mu m}$ (under a 100~V bias), we obtained $r_{\mathrm{eff,exp}} \approx 126~\mathrm{pm/V}$. The two independently obtained estimates are in good agreement, supporting the validity of the extracted effective Pockels coefficients.

 For the results in Fig.~4c, the two approaches yield estimates of approximately 147 and 156~pm/V, respectively, corresponding to an average value of $151~\mathrm{pm/V}$.

\subsection{High-speed characterization of the devices}

The electrical bandwidths (Fig.~4a) were characterized by measuring the electrical scattering parameter $S_{21,\mathrm{dB}} = 20 \log_{10}\left|\frac{V_{RF,\mathrm{Port2}}}{V_{RF,\mathrm{Port1}}}\right|$ using a vector network analyzer (P5003B, Keysight, 9~kHz--14~GHz). The VNA was connected to the device through two ground--signal--ground (GSG) probes (Picoprobe 40A series, GGB, DC--40~GHz) contacting the CPW pads (Fig.~2a) via 50~$\Omega$ SMA cables. Cable losses were de-embedded from the measured results.

The high-speed electro-optic characterization setup is shown in Fig.~4b. In addition to the instruments described above, a bias tee (ZX85-12G-S+, Mini-Circuits, 0.2~MHz--12~GHz) was used to combine the RF signal from VNA Port~1 (input power of 10~dBm) with a 7~V DC bias supplied by the Keithley 2400 SourceMeter. The combined signal was delivered to one GSG probe, while the opposite GSG probe was connected to a 50~$\Omega$ termination load. In the optical path, a tunable laser source (TSL-550, Santec) was operated at the wavelength corresponding to the maximum slope of the target resonance. The modulated optical transmission was amplified using an erbium-doped fiber amplifier (EDFA100S, Thorlabs), converted into an electrical signal using a high-speed photodetector (1544, New Focus, DC--12~GHz), and routed back to VNA Port~2. The measured $S_{21,\mathrm{dB}}$ responses therefore characterize the electro-optic bandwidths of the devices. The losses from the SMA cables and the bias tee were de-embedded from the final results.

% =====================
\section{Data availability}
All key data supporting the findings of this study are included in the main article and its Supplementary Information. Additional data sets and raw measurements are available from the corresponding author.

\section{Code availability}
The codes and simulation files that support the figures and data analysis in this article are available from the corresponding author.

\section{Acknowledgements}

The authors thank Joon-Suh Park, Pernille U. Fathi, and Jingchao Fang from Harvard University for the helpful discussions. The authors also thank Hana K. Warner, Houchen Li, and Yu-Fei Liu from Harvard University for their technical assistance. This research was supported by a Multidisciplinary University Research Initiative from the Air Force Office of Scientific Research (AFOSR Award No. FA9550-22-1-0307). This work was performed in part at the Harvard University Center for Nanoscale Systems (CNS); a member of the National Nanotechnology Coordinated Infrastructure Network (NNCI), which is supported by the National Science Foundation under NSF award no. ECCS-2025158. The simulation work was in part performed using Tidy3D software from Flexcompute Inc. M.O. acknowledges funding from the European Union (grant agreement 101076933 EUVORAM). The views and opinions expressed are, however, those of the author(s) only and do not necessarily reflect those of the European Union or the European Research Council Executive Agency. Neither the European Union nor the granting authority can be held responsible for them.

\section{Author contributions}

Conceptualization: Z.S., K.M.A.Y., M.O., A.A.D., and F.C. 

BTO growth and materials characterization: A.P., J.T., and A.A.D. 

Metasurface design and electromagnetic simulations: Z.S., K.M.A.Y., M.D., M.O., and I.B. 

Device fabrication: Z.S., K.M.A.Y., M.L.M., and T.P.L. 

Optical and electro-optic measurements: Z.S., K.M.A.Y., M.D., and X.L. 

RF design and high-speed characterization: Z.S., X.L., and A.S.-A. 

Data analysis and interpretation: Z.S. and M.D. 

Transmission electron microscopy characterization: Y.J., M.W., and X.P. 

Supervision: X.P., A.A.D., and F.C. 

Writing -- original draft: Z.S. 

Writing -- review \& editing: All authors.

\section{Competing interests}
The authors declare no competing interests.

% =====================
% References
% =====================
\bibliography{BTOref}

\end{document}

\begin{center}
{\Large \textbf{Supplementary Information}}\\[1em]
for\\[0.5em]
{\large Hybrid BaTiO$_3$/TiO$_2$ Metasurface for Efficient Gigahertz-Speed Free-Space Electro-Optic Modulation}
\end{center}

\section*{Supplementary Note 1: Characterization of the target TE guided-mode resonance at normal and oblique incidence}

This section describes simulated and experimental transmission spectra of the target transverse-electric (TE) guided-mode resonance (GMR) as a function of angle of incidence (AOI). First, we simulated the transmission response of an electrode-free, ideal two-dimensional array with the unit cell structure defined in Fig.~1b. The simulation was performed with rigorous coupled-wave analysis (RCWA), as detailed in the Methods. The results exhibit two resonance bands for the target GMR mode, as shown in Fig.~S2a.

These results can be explained by the phase-matching condition in GMRs, which estimates the resonance wavelength from in-plane momentum conservation between the in-plane wavevector of the incident field and the propagation constant of the guided mode modulo the grating reciprocal lattice vector~\cite{wang1993theory,641320}. For this case where the AOI changes along the $y$ direction, the associated reciprocal lattice vector is
\begin{equation}
G = \frac{2\pi}{\Lambda_y}.
\end{equation}
The phase-matching condition can then be written as
\begin{equation}
\beta(\omega) = k_{\parallel} + mG,
\label{eq:gmr_phase_matching}
\end{equation}
where $\beta(\omega)$ is the propagation constant of the guided mode, $m$ is the diffraction order, and
\begin{equation}
k_{\parallel} = k_0 n_{\mathrm{inc}} \sin\theta_{\mathrm{inc}}
\end{equation}
is the in-plane component of the incident wavevector. Here, $k_0 = 2\pi/\lambda$, $n_{\mathrm{inc}}$ is the refractive index of the incident medium, and $\theta_{\mathrm{inc}}$ is the angle of incidence.

Using $\beta(\omega) = k_0 n_{\mathrm{eff}}(\omega)$, where the effective
index is defined as a signed quantity,
\begin{equation}
n_{\mathrm{eff}}(\omega) \equiv \frac{\beta(\omega)}{k_0},
\end{equation}
Eq.~\eqref{eq:gmr_phase_matching} becomes
\begin{equation}
k_0 n_{\mathrm{eff}}(\omega)
=
k_0 n_{\mathrm{inc}}\sin\theta_{\mathrm{inc}}
+
m\frac{2\pi}{\Lambda_y}.
\end{equation}
Dividing both sides by $k_0 = 2\pi/\lambda$ gives the equivalent form
\begin{equation}
n_{\mathrm{eff}}(\omega)
=
n_{\mathrm{inc}}\sin\theta_{\mathrm{inc}}
+
m\frac{\lambda}{\Lambda_y}.
\end{equation}
Therefore, the GMR wavelength can be approximately estimated as
\begin{equation}
\lambda_{\mathrm{res}}
\approx
\frac{\Lambda_y}{m}
\left(
n_{\mathrm{eff}}(\omega) - n_{\mathrm{inc}}\sin\theta_{\mathrm{inc}}
\right).
\label{eq:gmr_wavelength_estimate}
\end{equation}
With this convention, counter-propagating guided modes have opposite signs of $n_{\mathrm{eff}}$.

This expression shows that when $\theta_{\mathrm{inc}}\neq 0$, the opposite diffraction orders $m$ and $-m$ couple to counter-propagating guided modes at different wavelengths, which qualitatively explains why the target GMR is split into two bands at oblique incidences. However, the simple phase-matching picture does not give exact resonance wavelengths. For example, Eq.~\eqref{eq:gmr_wavelength_estimate} would predict the two counter-propagating modes to be degenerate at the same wavelength $\lambda_{\mathrm{res}}\approx
\Lambda_y |n_{\mathrm{eff}}|/|m|$ for normal incidence. However, the two bands in Fig.~S2a do not merge into one resonance when $\theta_{\mathrm{inc}} = 0$. Instead, the two bands remain spectrally separated near
$\theta_{\mathrm{inc}}=0$, forming an anti-crossing. This phenomenon results from the Bragg scattering between the two originally degenerate counter-propagating guided modes with propagation constants $+\beta$ and $-\beta$, so that the eigenmodes under this condition are standing-wave-like supermodes formed by their symmetric and antisymmetric
linear combinations~\cite{kogelnik1969coupled,yariv1982optical}. 

Under the $180^\circ$ rotational symmetry with respect to the $z$ axis ($C_2$ symmetry), the supermode on the longer-wavelength band is an odd mode, as confirmed by the $x$--$y$ plane $\mathrm{Re}(E_x)$ field profile shown in Fig.~S2b(i), while the one on the shorter-wavelength band is an even mode. Since plane waves at normal incidence are odd under $C_2$, the coupling from the incident wave to the even mode is forbidden, which causes the
shorter-wavelength branch to disappear from the normal-incidence transmission spectrum. The corresponding dark eigenmode is a symmetry-protected bound state in the continuum (SP-BIC)~\cite{hsu2016bound}. At nonzero AOI, the excitation carries finite in-plane momentum, so the symmetry protection is lifted and the originally dark BIC mode can couple to radiation, resulting in a high-$Q$ quasi-BIC resonance. The $\mathrm{Re}(E_x)$ field profile of this quasi-BIC mode is shown in Fig.~S2b(ii). Note that this mode remains predominantly even under $C_2$.

The experimental characterization of the transmission spectra at different AOIs along the $y$ direction is performed using the device with a $2~\mathrm{mm} \times 2~\mathrm{mm}$ metasurface described in the main article. As shown in Fig.~S2c, the measured transmittance map is in good agreement with the simulated one in terms of the trend of the two resonance bands. The transmittance spectra in Fig.~S2d show only the odd mode appearing as a visible resonance at $\mathrm{AOI}=0^\circ$ and two separated resonances at $0.5^\circ$ and $1^\circ$, further confirming the theory described above. In spite of this, there are two discrepancies between the simulations and the measurements. First, the resonance contrast is lower in the measurement because the tested device contains gold interdigitated electrodes (IDEs), unlike the electrode-free array in the simulation. The electrodes can reduce the contrast through several channels: additional optical loss that raises the intrinsic loss rate, a non-resonant transmission background over the fraction of the area they cover, and lateral truncation of the guided mode within each electrode gap. Second, the resonances at oblique incidences in the measurement have lower quality factors ($Q$ factors) than those in the simulation, possibly because the random scattering loss due to fabrication defects sets an upper bound for the $Q$ factor.  

We further measured the transmission spectra under applied DC voltages at different AOIs along the $y$ direction, as shown in Fig.~S3. The blueshift of the resonance at normal incidence (Fig.~S3a) is also shown in the main article. As the AOI increases, the spectral separation between the two resonances increases. However, all seven resonances shown exhibit a similar wavelength shift rate with increasing voltage. This phenomenon can be explained by perturbation theory, which gives the approximate resonance wavelength shift relation~\cite{Benea2022EOPolymer,Soma2025EOPolymer,PhCBook}
\begin{equation}
\frac{\Delta \lambda_{\mathrm{res}}}{\lambda_{\mathrm{res}}}
\approx
\Gamma_{\mathrm{AR}} \frac{\Delta n_{\mathrm{BTO}}}{n_{\mathrm{BTO}}},
\label{eq:Shift}
\end{equation}
where $\Gamma_{\mathrm{AR}}$ denotes the optical confinement factor in the active region, and $\Delta n_{\mathrm{BTO}} = -\tfrac{1}{2} n_{\mathrm{BTO}}^{3} r_{\mathrm{eff}} E_{\mathrm{DC}}$. For small AOIs, $\lambda_{\mathrm{res}}$, $\Gamma_{\mathrm{AR}}$, $r_{\mathrm{eff}}$, and $n_{\mathrm{BTO}}$ are only weakly perturbed, leading to similar values of $\Delta\lambda_{\mathrm{res}}/E_{\mathrm{DC}}$. Because $E_{\mathrm{DC}}$ scales linearly with the applied voltage, these resonances exhibit similar voltage tuning rates in Fig.~S3.

\section*{Supplementary Note 2: Fabrication process of hybrid BTO/TiO$_2$ metasurfaces}

The fabrication process of the hybrid BTO/TiO$_2$ metasurfaces is illustrated in Fig.~S4. The process consists of two main steps: fabrication of the gold electrodes and alignment markers and fabrication of the aligned metasurface.

First, gold electrodes and alignment markers were fabricated on our BTO sample using a lift-off process. A 600-nm-thick ZEP520A (Zeon SMI) layer was spin-coated on the sample and patterned by electron-beam lithography (EBL, Elionix Boden-150). The IDE pattern was oriented at a $45^\circ$ angle with respect to the diced chip edge. After development, a gold layer with a thickness of 240~nm (for the $2~\mathrm{mm} \times 2~\mathrm{mm}$ device in Fig.~3) or 55~nm (for devices studied in Fig.~4) was deposited by electron-beam evaporation (Denton E-Beam Evaporator or PVD Products E-Beam Deposition System). The residual resist was then removed by Remover PG, leaving the patterned gold electrodes and markers on the sample surface, as shown in Fig.~S4a.

Next, the TiO$_2$ metasurface was fabricated using an aligned Damascene process, as shown in Fig.~S4b. A 200-nm-thick layer of 1:1 diluted ZEP520A was spin-coated on the sample. After being aligned to the previously defined gold markers, the metasurface pattern was exposed by EBL and developed to form the resist template. TiO$_2$ was conformally deposited into the patterned resist openings at $90\,^{\circ}\mathrm{C}$ by atomic layer deposition (ALD, Cambridge Nanotech Savannah). After deposition, the excess TiO$_2$ on top of the resist was removed by an etch-back process, leaving TiO$_2$ only inside the patterned regions. Finally, the remaining resist was removed by dry ashing, resulting in the hybrid BTO/TiO$_2$ metasurface structure.

This process enables accurate alignment of the metasurface pattern with respect to the gold electrodes while avoiding direct etching of the BTO layer. The use of the Damascene process also allows the TiO$_2$ nanostructures to be defined by deposition and planarization rather than by subtractive etching, which helps preserve the underlying BTO film.

\section*{Supplementary Note 3: Effects of the electrode thickness}

The Au electrodes of the $2~\mathrm{mm} \times 2~\mathrm{mm}$ device studied in Fig.~3 are $240~\mathrm{nm}$ thick, whereas those of the devices studied in Fig.~4 are $55~\mathrm{nm}$ thick. To verify that this difference does not significantly affect either the static or optical fields in the active region, we repeated the electrostatic and FDTD optical simulations using an Au thickness of $t_{\mathrm{Au}} = 55~\mathrm{nm}$. The results are shown in Figs.~S5 and S6, respectively.

The electrostatic simulations show that the static electric field within the active region for $t_{\mathrm{Au}} = 55~\mathrm{nm}$ is nearly identical to that for $t_{\mathrm{Au}} = 240~\mathrm{nm}$. For $t_{\mathrm{Au}} = 55~\mathrm{nm}$, the field remains highly uniform, varying by less than $0.6\%$, and is aligned predominantly along the $x$ direction, with the spatially averaged squared magnitudes of $E_y$ and $E_z$ on the order of $\sim 10^{-6}$ of that of $E_x$. As an example, the $E_x$ fields along the line of intersection between the $x$--$y$ mid-plane of the BTO layer and the $x$--$z$ plane through the centers of the TiO$_2$ nanopillars are shown in Fig.~S5a,b. At any point within the active region along this line, the relative difference between the fields for $t_{\mathrm{Au}} = 240$ and $55~\mathrm{nm}$ is less than $0.03\%$. The field magnitude averaged over the whole active region, $\langle |E_{\mathrm{DC}}| \rangle$, as a function of $t_{\mathrm{Au}}$ is shown in Fig.~S5c. The relative difference in $\langle |E_{\mathrm{DC}}| \rangle$ between $t_{\mathrm{Au}} = 55$ and $240~\mathrm{nm}$ is only $-0.018\%$. This negligible difference results from the fact that the gap field is primarily determined by the electrode separation and the high permittivity of the BTO film, rather than by the electrode cross-section. In addition, the edge-fringing perturbation decays within $\sim t_{\mathrm{Au}}$ of the electrode, well outside the active region. Since the extracted $r_{\mathrm{eff,exp}}$ scales inversely with $E_{\mathrm{DC}}$, the corresponding correction is below $0.02\%$.

The optical mode is likewise insensitive to $t_{\mathrm{Au}}$. For different $t_{\mathrm{Au}}$, the transmission spectra are nearly identical (Fig.~S6a). Under increasing voltage, the resonances for $t_{\mathrm{Au}} = 240$ and $55~\mathrm{nm}$ blueshift at essentially the same shift rate $S_{\mathrm{sim}}$ (Fig.~S6b,c). The optical field profiles at resonance also show negligible differences between the two Au thicknesses. As shown in Fig.~S6d, the squared electric-field magnitudes, $|E|^2$, exhibit nearly identical profiles along the same line of intersection defined in the electrostatic simulations. The relative difference in $|E|^2$ in the $x$--$z$ plane through the centers of the TiO$_2$ nanopillars is below $\sim 1\%$ between $t_{\mathrm{Au}} = 240$ and $55~\mathrm{nm}$, as shown in Fig.~S6e. The relative root mean square error (RMSE) in $|E|^2$ between the two Au thicknesses is only $\sim 0.28\%$ when evaluated over the whole plane and $\sim 0.17\%$ when evaluated over the active region in this plane. These results indicate nearly identical modal profiles for $t_{\mathrm{Au}} = 240$ and $55~\mathrm{nm}$ and, therefore, very similar field confinement factors $\Gamma_{\mathrm{AR}}$. This insensitivity is expected: the optical field decays within a few tens of nanometers in gold at telecommunication wavelengths, so the mode already experiences an optically thick electrode at $t_{\mathrm{Au}} = 55~\mathrm{nm}$.

In conclusion, the electrode-thickness difference cannot account for the $\sim 25\%$ device-to-device variation in $r_{\mathrm{eff,exp}}$.

\section*{Supplementary Note 4: Analysis of the $r_{42}$-mediated Pockels effect in orthogonal-domain $a$-oriented BTO}

This section explains why the device is oriented along the diagonal direction and analyzes the ferroelectric hysteresis measurements shown in Fig.~3f. In this note, we focus on the contribution from the $r_{42}$ coefficient because $r_{42}$ is the largest Pockels tensor element in BTO, as mentioned in the main article, and therefore can provide the dominant Pockels response when it is efficiently activated by the device geometry. The same general framework can be extended to other electro-optic tensor elements by evaluating their contributions in the local crystallographic coordinate system of each domain.

As illustrated in Fig.~S7a, the thin-film BTO used in this work contains two types of mutually orthogonal $a$-oriented ferroelectric domains. The crystallographic $c$ axis lies in the device plane and is oriented along either the $x$ or $y$ direction, corresponding to Domain~1 and Domain~2, respectively. Note that the spatial coordinate system defined in Fig.~1a in the main text does not apply to this section.

In the local crystallographic coordinate system of either domain, the Pockels effect is described by the field-induced perturbation of the optical impermeability tensor,
\begin{equation}
\Delta \eta_{ij}
=
\Delta \left(\frac{1}{n^2}\right)_{ij},
\end{equation}
where $\eta_{ij}$ is the impermeability tensor. In contracted notation, the Pockels effect can be written as~\cite{BoydNonlinear}
\begin{equation}
\Delta \eta_i = \sum_{j=1}^{3} r_{ij} E_j,
\label{eq:pockels_contracted}
\end{equation}
where $i=1,\ldots,6$ denotes the contracted tensor components
\begin{equation}
1\rightarrow XX,\quad
2\rightarrow YY,\quad
3\rightarrow ZZ,\quad
4\rightarrow YZ,\quad
5\rightarrow XZ,\quad
6\rightarrow XY,
\end{equation}
and $X,Y,Z$ denote the local crystallographic axes of BTO, with $Z$ chosen along the $c$ axis.

For tetragonal BTO with $4mm$ symmetry, the Pockels tensor can be written as
\begin{equation}
\begin{pmatrix}
\Delta \eta_1\\
\Delta \eta_2\\
\Delta \eta_3\\
\Delta \eta_4\\
\Delta \eta_5\\
\Delta \eta_6
\end{pmatrix}
=
\begin{pmatrix}
0 & 0 & r_{13}\\
0 & 0 & r_{13}\\
0 & 0 & r_{33}\\
0 & r_{42} & 0\\
r_{42} & 0 & 0\\
0 & 0 & 0
\end{pmatrix}
\begin{pmatrix}
E_X\\
E_Y\\
E_Z
\end{pmatrix}.
\label{eq:bto_r_matrix}
\end{equation}
Here, $E_X$, $E_Y$, and $E_Z$ are the components of the applied static electric field in the local crystallographic coordinate system. The diagonal coefficients $r_{13}$ and $r_{33}$ modify the principal values of the refractive-index ellipsoid. In contrast, the coefficient $r_{42}$ produces off-diagonal perturbations of the impermeability tensor,
\begin{equation}
\Delta \eta_4 = r_{42}E_Y,
\qquad
\Delta \eta_5 = r_{42}E_X.
\end{equation}
Equivalently, in full tensor notation,
\begin{equation}
\Delta \boldsymbol{\eta}_{r_{42}}
=
\frac{r_{42}}{2}
\begin{pmatrix}
0 & 0 & E_X\\
0 & 0 & E_Y\\
E_X & E_Y & 0
\end{pmatrix}.
\label{eq:r42_full_tensor}
\end{equation}
Thus, the $r_{42}$ term couples the $Z$ axis to the transverse $X$ and $Y$ axes. This off-diagonal perturbation rotates the index ellipsoid and changes the refractive index experienced by an optical field whose polarization has projections along both the $c$ axis and the transverse axes.

For an optical field polarized along a unit vector
\begin{equation}
\hat{\mathbf{u}} = u_X\hat{\mathbf{X}} + u_Y\hat{\mathbf{Y}} + u_Z\hat{\mathbf{Z}},
\end{equation}
the $r_{42}$-mediated perturbation to the impermeability is
\begin{equation}
\Delta\left(\frac{1}{n_u^2}\right)_{r_{42}}
=
\hat{\mathbf{u}}^{T}
\Delta \boldsymbol{\eta}_{r_{42}}
\hat{\mathbf{u}}
=
r_{42}\left(u_Xu_ZE_X + u_Yu_ZE_Y\right).
\label{eq:r42_delta_inverse_n2}
\end{equation}
For a small perturbation, this corresponds to an approximate refractive-index change
\begin{equation}
\Delta n_u
\approx
-\frac{1}{2}n_u^3
r_{42}\left(u_Xu_ZE_X + u_Yu_ZE_Y\right).
\label{eq:r42_delta_n}
\end{equation}
Eq.~\eqref{eq:r42_delta_n} shows that the $r_{42}$ effect depends on the relative orientation between the optical polarization, the static electric field, and the local crystallographic axes.

Eq.~\eqref{eq:r42_delta_n} explains why both the static electric field and the optical polarization are chosen to be oriented at $45^\circ$ with respect to the $c$ axes, as illustrated in Fig.~S7b,c. For an in-plane field polarized at an angle $\phi$ with respect to the local $c$ axis, the refractive-index change is proportional to 
\begin{equation} 
u_{\perp}u_Z = \sin\phi\cos\phi = \frac{1}{2}\sin 2\phi, 
\end{equation} 
which is maximized at $\phi=45^\circ$. Thus, a $45^\circ$ in-plane optical polarization is optimal for both Domain~1 and Domain~2.

The same conclusion also applies to the static field under the assumption that the populations of Domain~1 and Domain~2 are equal. To see this, we write the in-plane DC field as $\mathbf{E}_{\mathrm{DC}}=E_{\mathrm{DC}}(\cos\phi\,\hat{\mathbf{x}}+\sin\phi\,\hat{\mathbf{y}})$. The transverse field component that activates
the $r_{42}$ response scales as $E_{\mathrm{DC}}\sin\phi$ in Domain~1 and $E_{\mathrm{DC}}\cos\phi$ in Domain~2. If the two domain types have equal area fractions, their averaged response is therefore proportional to
\begin{equation}
\frac{1}{2}\left(\cos\phi+\sin\phi\right)
=\frac{1}{\sqrt{2}}\cos(\phi-45^\circ),
\end{equation}
which is maximized at $\phi=45^\circ$. Thus, orienting the static field at $45^\circ$ balances the field projections onto the two orthogonal domain types and maximizes the $r_{42}$-mediated Pockels response.

To explain the ferroelectric hysteresis observed in Fig.~3f, we consider the elementary polarization-reversal process illustrated in Fig.~S8. Starting from the initial configuration in Fig.~S8a, the static field induces an
$r_{42}$-mediated perturbation of the refractive-index ellipse, reducing the refractive index experienced by the indicated optical polarization. When the external static field is reversed, the ferroelectric polarization does not
immediately switch because of hysteresis. This produces the intermediate configuration shown in Fig.~S8b, where the
$r_{42}$-induced refractive-index perturbation changes sign and the refractive index for the indicated optical polarization is increased relative to the unperturbed value. As the magnitude of the reversed field further increases, the ferroelectric polarization switches. The system then enters the configuration shown in Fig.~S8c, where the sign of the $r_{42}$ perturbation is restored to that in the initial state and the refractive index is again reduced relative to the unperturbed value.

To explain the collective behavior quantitatively, the configurations in Fig.~S8a,c are defined as the $u$ state, whereas the intermediate configuration in Fig.~S8b is defined as the $v$ state. For the indicated optical polarization, the $r_{42}$-mediated refractive-index perturbations in the two states have opposite signs. We therefore write
\begin{equation}
\Delta n_u = -\Delta n_0, \qquad
\Delta n_v = \Delta n_0,
\end{equation}
where $\Delta n_0$ is the magnitude of the index perturbation given by Eq.~\eqref{eq:r42_delta_n} and can be simplified as $\Delta n_0 = \tfrac{1}{2} n_0^{3} |r_{\mathrm{eff}} E_{\mathrm{DC}}|$ in this discussion. Although the discussion above is based on Domain~2, the same two states can be generalized to Domain~1 since the static field is oriented at $45^\circ$ with respect to the two orthogonal in-plane $c$-axis orientations. With both domains taken into account, we denote the area-weighted fractions of the $u$ and $v$
states by $U$ and $V$, respectively, with
\begin{equation}
U+V=1.
\end{equation}
The effective refractive-index perturbation experienced by the optical mode
can then be approximated as
\begin{equation}
\Delta n_{\mathrm{eff}}
\approx
U\Delta n_u+V\Delta n_v
=
(V-U)\Delta n_0
=
\tfrac{1}{2} (V-U) n_0^{3} |r_{\mathrm{eff}} E_{\mathrm{DC}}|.
\label{eq:uv_delta_n}
\end{equation}
Thus, the sign and magnitude of the effective electro-optic response are determined by the imbalance between the $u$-state and $v$-state populations.

Now we consider the reverse sweep in Fig.~3f. Based on Eq.~\eqref{eq:Shift}, the wavelength shift $\Delta \lambda_{\mathrm{res}}$ is approximately proportional to $\Delta n_{\mathrm{eff}}$ for small wavelength shifts. After poling, the populations $U\approx1$ and $V\approx0$ can be assumed, so the effective refractive-index perturbation becomes $\Delta n_{\mathrm{eff}}=-\tfrac{1}{2}n_0^{3} |r_{\mathrm{eff}} E_{\mathrm{DC}}|$. The approximately linear trend between the resonance wavelength and voltage from 100~V to 0~V in Fig.~3f suggests that the populations remain almost unchanged at this stage. When the voltage polarity is reversed, the static field changes sign while the ferroelectric polarization initially remains unchanged. Therefore, the regions that were in the $u$ response configuration enter the $v$ response configuration, giving $U\approx0$ and $V\approx1$. As the magnitude of the negative voltage increases, the ferroelectric domains begin to switch, and $V-U$ starts to decrease from 1, resulting in a reduced wavelength-shift rate. The resonance wavelength reaches its maximum near the switching midpoint, where $V$ and $U$ become comparable. The field associated with this maximum provides an optical estimate of the coercive field. After the switching midpoint, as the magnitude of the negative voltage continues to increase, $V-U$ decreases from 0 to $-1$, restoring the fully poled state with the opposite ferroelectric polarization direction.

\section*{Supplementary Note 5: Origin of the reduced Fano lineshape contrast in the experiments in Fig.~3c}

The guided-mode resonance exhibits a Fano lineshape because the transmitted field is the coherent superposition of a direct transmission pathway and radiation resonantly scattered from the guided mode, as illustrated in Fig.~S9a. The transmittance can be written as~\cite{Fan2002GMR}
\begin{equation}
T(\omega)
=
\left|
t
\right|^2
=
\left|
t_\mathrm{d}
+
t_{\mathrm{res}}
\right|^2
=
\left|
t_\mathrm{d}
+
\frac{a}{i(\omega-\omega_0)+\gamma}
\right|^2,
\label{eq:fano_transmission}
\end{equation}
where $t$ is the total transmission amplitude, $t_\mathrm{d}$ is the direct transmission amplitude, $t_\mathrm{res}$ is the resonant transmission amplitude, $a$ is the complex resonance amplitude coefficient, $\omega_0$ is the resonance angular frequency, and $\gamma$ is the resonance linewidth parameter. An illustrative example of the Fano resonance is shown in Fig.~S9b, where the asymmetric Fano lineshape originates from coherent interference between the direct and resonant transmission pathways. To produce a pronounced asymmetric Fano feature, the coherence between these two pathways must be maintained.

In experiments, the Fano asymmetry is generally weaker than that in ideal full-wave simulations. This reduction can arise from several physical mechanisms. First, material absorption and scattering loss increase $\gamma$ and reduce $t_\mathrm{res}$. This broadens the resonance, lowers the quality factor, and suppresses the Fano feature. For our devices, we performed simulations with additional artificial extinction coefficients in BTO and TiO$_2$ to emulate absorption-induced loss, as shown in Fig.~S9c. Increasing the extinction coefficient reduces the $Q$ factor and transmission contrast and suppresses the Fano feature. This confirms that the residual loss can contribute to the reduced Fano asymmetry in our devices.

Second, spatial inhomogeneity in the device, such as variations in the angle of incidence, BTO thickness, TiO$_2$ geometry, or local refractive index, causes different regions of the metasurface to resonate at slightly different frequencies. The measured spectrum is then an average over a distribution of resonance parameters, including $a$, $\omega_0$, and $\gamma$, which further broadens the resonance and reduces the apparent Fano asymmetry.

Third, incoherent direct transmission can suppress the Fano lineshape without substantially changing the resonance linewidth. This incoherent direct transmission does not preserve a well-defined phase relationship with the resonantly scattered field and therefore does not participate in the coherent interference described by Eq.~\eqref{eq:fano_transmission}. Instead, it contributes as an incoherent background intensity, which reduces the Fano asymmetry. In the experiment, surface roughness and fabrication imperfections can scatter part of the transmitted light into diffuse background pathways, reducing the coherent phase information associated with the layer thickness and metasurface geometry.

\section*{Supplementary Note 6: Fano fits and extraction of the modulation efficiency $\eta_\mathrm{mod}$}

The normalized transmittance spectra were fitted using a phenomenological Fano lineshape model to extract the resonance wavelength, linewidth, and quality factor. For each spectrum, a local fitting window centered around the transmittance minimum was selected to isolate the resonance feature. Within this window, the transmittance was modeled as a slowly varying linear background plus a Fano resonance term,
\begin{equation}
T(\lambda)=B_0+B_1(\lambda-\lambda_0)-D F(\lambda),
\end{equation}
where
\begin{equation}
F(\lambda)=1-\frac{(q+\epsilon)^2}{(1+q^2)(1+\epsilon^2)},
\qquad
\epsilon=\frac{2(\lambda-\lambda_0)}{\Gamma}.
\end{equation}

Here, $B_0$ and $B_1$ describe the local background, $D$ is the resonance contrast parameter, $q$ is the Fano asymmetry parameter, $\lambda_0$ is the fitted Fano resonance wavelength parameter, and $\Gamma$ is the linewidth parameter. The fitting quality was evaluated using the root-mean-square error and coefficient of determination $R^2$ between the fitted curve and the data within the local fitting window. The resonance quality factor was estimated from the fitted parameters as $Q = \lambda_0/\Gamma$. The covariance matrix of the fitted parameters was estimated from the fitting Jacobian and the residual mean-square error. The standard error of $Q=\lambda_0/\Gamma$ was then calculated using first-order error propagation, including the covariance between $\lambda_0$ and $\Gamma$.

The modulation efficiency $\eta_\mathrm{mod}$ of the $2~\mathrm{mm} \times 2~\mathrm{mm}$ device is extracted by directly evaluating the transmittance under different applied voltages at a fixed operating wavelength (Fig.~3e), as described in the main article. However, due to the relatively large spectral fluctuations in the measurements of the $0.3~\mathrm{mm} \times 0.3~\mathrm{mm}$ device, the method above does not work well. Instead, the modulation efficiency $\eta_\mathrm{mod}$ of the $0.3~\mathrm{mm} \times 0.3~\mathrm{mm}$ device was estimated from the measured resonance shift rate and the maximum slope of the resonance spectrum extracted from the Fano fitted curve. For small voltage-induced perturbations, the modulation efficiency can be expressed through the chain rule as

\begin{equation}
\frac{\mathrm{d}T}{\mathrm{d}V}
=
\frac{\mathrm{d}T}{\mathrm{d}\lambda}
\frac{\mathrm{d}\lambda}{\mathrm{d}V},
\end{equation}

where $T$ is the normalized transmittance, $\lambda$ is the optical wavelength, and $V$ is the applied voltage. The maximum resonance slope, $\mathrm{d}T/\mathrm{d}\lambda = 0.303~\mathrm{nm}^{-1}$, was obtained from the resonance flank of the Fano fit at 0~V (see Fig.~S10b). The resonance shift rate, $\mathrm{d}\lambda/\mathrm{d}V = 0.065~\mathrm{nm/V}$, was extracted from the measured shift of the resonance wavelength over 0 to 20~V, a low bias range chosen to reflect practical device operating conditions. The modulation efficiency, $\eta_\mathrm{mod}=\mathrm{d}T/\mathrm{d}V=0.020~\mathrm{V^{-1}}$, was then calculated as the product of these two independently measured quantities. Note that the shift rate quoted in the main article, $S_{\mathrm{exp}} \approx 0.071~\mathrm{nm/V}$, is instead evaluated at the maximum applied voltage of $100~\mathrm{V}$, the condition closest to full poling of the BTO and therefore the appropriate one for extracting the effective Pockels coefficient.

\section*{Supplementary Note 7: Analysis of the factors limiting the EO modulation bandwidths}

The setups for the electrical and electro-optic bandwidth measurements are shown in Fig.~S11a,b, respectively. In addition to the electrical configuration and EO configuration~\textit{A} discussed in the main text, we also measured the high-speed modulation response of EO configuration~\textit{B}, in which the bottom RF probe is lifted off from the device contact pads, effectively leaving the RF circuit unterminated. In all configurations, the components in the RF pathways are connected using $50~\Omega$ SMA cables. Although these cables should be treated as transmission lines at GHz frequencies, the devices themselves are electrically small compared with the RF wavelength over the frequency range considered in this study; for example, $\lambda_{\mathrm{RF}} \approx 12~\mathrm{cm}$ at $1~\mathrm{GHz}$. Therefore, the device response can be modeled using a lumped-element approximation. Although impedance mismatch between the lumped device and the $50~\Omega$ RF environment can produce reflections and frequency-dependent ripples, the monotonic bandwidth roll-off is primarily captured by the lumped RC response of the device capacitance loaded by the relevant impedances.

First, we estimate the total capacitance and resistance of the $0.3~\mathrm{mm} \times 0.3~\mathrm{mm}$ device under a lumped-element approximation. Because the permittivity of BTO is much larger than those of the other materials in our devices, we only consider the electric-field flux through the BTO layer and estimate the capacitance per unit length of a single gap as
\begin{equation}
\begin{aligned}
    C' 
    &\approx \varepsilon_0 \varepsilon_x \frac{t_{\mathrm{BTO}}}{g} \\
    &\approx 
    \left(8.85 \times 10^{-12}~\mathrm{F/m}\right)
    \times 1044
    \times \frac{0.28~\mu\mathrm{m}}{7~\mu\mathrm{m}} \\
    &\approx 0.37~\mathrm{nF/m},
\end{aligned}
\label{eq:Cprime}
\end{equation}
where $\varepsilon_0$ is the vacuum permittivity, $\varepsilon_x$ is the relative in-plane permittivity of BTO (see the Electrostatic simulation section in Methods), $t_{\mathrm{BTO}}$ is the thickness of the BTO layer, and $g$ is the gap distance between two adjacent IDEs. The $0.3~\mathrm{mm}$ IDE length gives a capacitance of $0.11~\mathrm{pF}$ per gap. The device contains 39 gaps, corresponding to 39 capacitors in parallel and resulting in a total capacitance of $C \approx 4.33~\mathrm{pF}$. The resistance of one IDE finger is estimated using the standard resistivity relation
\begin{equation}
\begin{aligned}
    R_{\mathrm{IDE}} 
    &\approx \rho_{\mathrm{Au}}\frac{L}{w_{\mathrm{Au}} t_{\mathrm{Au}}} \\
    &\approx \left(2.44 \times 10^{-8}~\Omega\cdot\mathrm{m}\right)
    \frac{3 \times 10^{-4}~\mathrm{m}}
    {\left(1 \times 10^{-6}~\mathrm{m}\right)
    \left(55 \times 10^{-9}~\mathrm{m}\right)} \\
    &\approx 133~\Omega,
\end{aligned}
\end{equation}
where $\rho_{\mathrm{Au}}$ is the resistivity of Au, $L$ is the IDE length, $w_{\mathrm{Au}}$ is the IDE width, and $t_{\mathrm{Au}}$ is the Au electrode thickness. The device contains 40 IDE fingers. As a lumped estimate, we approximate these fingers as 40 resistors in parallel, giving an effective device resistance of $R \approx 3.33~\Omega$. This is small compared with the $25~\Omega$ presented by the $50~\Omega$ source and termination in parallel, so the RC time constant is dominated by the device capacitance charged through the external $50~\Omega$ environment. Since the finger resistance scales as $1/t_{\mathrm{Au}}$, it would become comparable to the $50~\Omega$ environment only for thicknesses well below the $55~\mathrm{nm}$ used here. The device resistance therefore contributes only a small correction and is neglected in the following analysis.

The lumped-element RF equivalent circuit models are shown in Fig.~S11c, in which the DC-bias branch of the bias tee is omitted for clarity. The models include the $50~\Omega$ source resistance of the vector network analyzer (VNA), the $4.33~\mathrm{pF}$ device capacitance, and, when present, the $50~\Omega$ termination resistance. The top circuit in Fig.~S11c corresponds to the electrical configuration and EO configuration~\textit{A}, in which the termination resistance is present. The Thevenin resistance seen by the device capacitance is therefore the parallel combination of the two $50~\Omega$ resistances, giving $R_{\mathrm{A}} = 25~\Omega$. The corresponding theoretical $-3~\mathrm{dB}$ bandwidth is
\begin{equation}
    f_{\mathrm{A},-3\,\mathrm{dB}}
    = \frac{1}{2\pi R_{\mathrm{A}}C}
    \approx 1.47~\mathrm{GHz}.
\end{equation}
For the bottom, unterminated circuit, the Thevenin resistance is simply $R_{\mathrm{B}} = 50~\Omega$, resulting in a theoretical bandwidth of
\begin{equation}
    f_{\mathrm{B},-3\,\mathrm{dB}}
    = \frac{1}{2\pi R_{\mathrm{B}}C}
    \approx 0.74~\mathrm{GHz},
\end{equation}
which is half of the bandwidth of the top circuit. This difference is confirmed by the measurement results in Fig.~S11d: EO configuration~\textit{B} has the lowest electro-optic bandwidth ($\sim 0.6~\mathrm{GHz}$), whereas the electrical configuration and EO configuration~\textit{A} reach bandwidths of $\sim 1~\mathrm{GHz}$ and $\sim 0.8~\mathrm{GHz}$, respectively. The measured electrical bandwidth ($\sim 1~\mathrm{GHz}$) is lower than the ideal RC estimate ($1.47~\mathrm{GHz}$), likely because the lumped model omits nonideal parasitics such as probe/contact parasitics. The IDE resistance estimated above accounts for only a small part of this discrepancy.

Another discrepancy in the experimental results is that the bandwidth of EO configuration~\textit{A} is slightly lower than the electrical bandwidth, although their equivalent RF circuits are the same. In the main text, we attribute this discrepancy to RF-frequency dispersion of $r_{\mathrm{eff}}$, including the dispersion inherited from $\chi^{(1)}$ through Miller's rule and a reduced acoustic/strain-mediated contribution. The latter mechanism can contribute significantly to $r_{\mathrm{eff}}$ under DC bias, but its contribution can be dynamically clamped at high frequencies because the acoustic deformation cannot fully follow the RF drive. The resonance-like signatures marked by the two arrows in Fig.~S11d provide evidence for an acoustic/strain-mediated contribution. Because these signatures robustly appear in both EO responses but are absent from the electrical response, they are likely associated with acoustic resonances in the device rather than RF-circuit artifacts. We also note that the high-speed photodetector (DC--12~GHz), which is present only in the EO configurations and is not de-embedded in the final results, is a less likely origin of the discrepancy in the roll-offs between the electrical response and the EO configuration~\textit{A} response.

We note that although the terminated circuit provides a higher bandwidth, it comes at the expense of modulation efficiency. In Fig.~S11c (top), the source RF voltage is divided between the source and termination resistances, so the voltage applied to the device is only half of the source voltage. In contrast, in the unterminated circuit, nearly the full source voltage is applied to the device. Moreover, for our specific EO configurations, the terminated circuit introduces an additional static power dissipation of $\sim 1~\mathrm{W}$ because the $7~\mathrm{V}$ DC bias is applied across the $50~\Omega$ termination resistance. However, this static power consumption is not intrinsic to RF termination and can be avoided by AC-coupling the termination, for example by connecting the termination resistor in series with a large DC-blocking capacitor.

Finally, we estimate the potential of our platform for achieving a $10~\mathrm{GHz}$ bandwidth without sacrificing much of the modulation efficiency. If the metasurface size is reduced to $0.1~\mathrm{mm} \times 0.1~\mathrm{mm}$ while maintaining the current supercell structure, the device capacitance would decrease to $1/9$ of the current value, whereas the device resistance would remain nearly unchanged and can still be neglected according to the estimates above. Therefore, the ideal RC-limited bandwidth would simply increase by a factor of 9, giving an ideal electrical bandwidth of $\sim 13.23~\mathrm{GHz}$. The main trade-off is the smaller device aperture, while the modulation efficiency could be mostly retained.

\FloatBarrier

\begin{figure}
\centering
\includegraphics[width=0.9\linewidth]{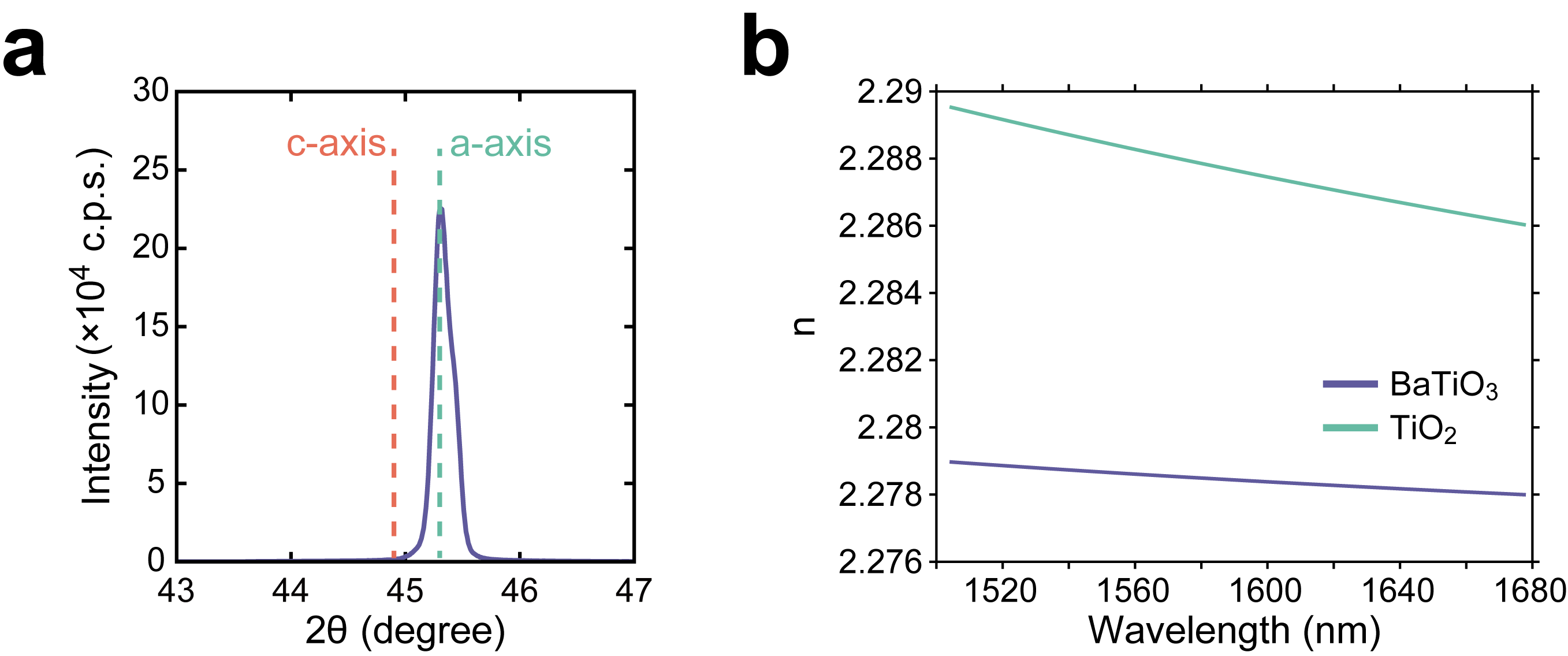}% Here is how to import EPS art
\caption{\textbf{Material characterizations.} \textbf{a}, X-ray diffraction (XRD) scan of the RF magnetron-sputtered BTO film. The $2\theta$ positions corresponding to the $a$- and $c$-axis lattice constants are indicated. The measured peak coincides with the $a$-axis position, indicating that the BTO $c$-axis (polar axis) lies in-plane. \textbf{b}, Refractive indices of BTO and TiO$_2$ measured by ellipsometry over the 1500--1680 nm wavelength range. The optical anisotropy of BTO was not accounted for in the ellipsometry fitting, and an isotropic model was assumed.}
\label{fig:figS1}
\end{figure}

\begin{figure}
\centering
\includegraphics[width=1\linewidth]{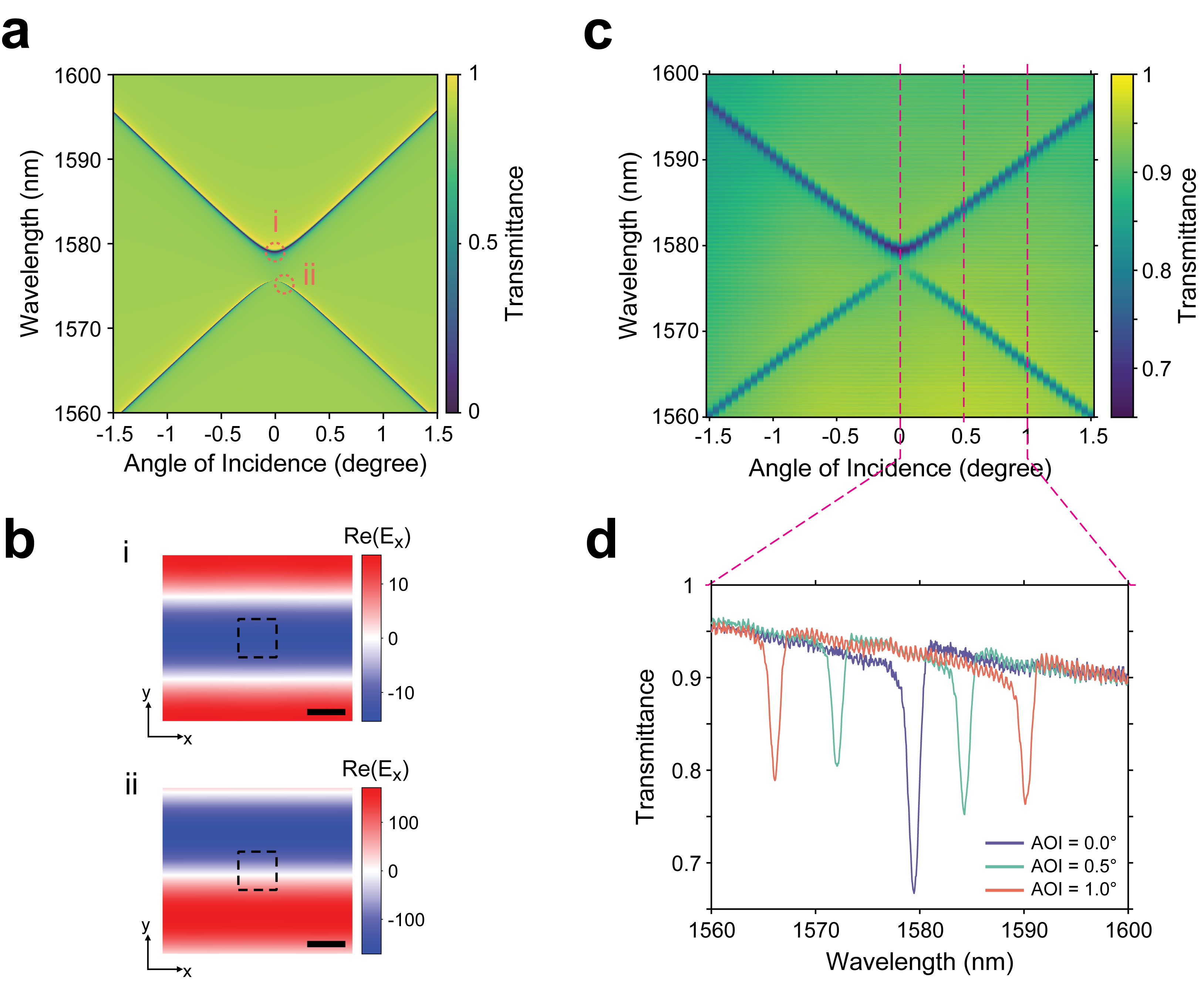}% Here is how to import EPS art
\caption{\textbf{Characterization of the target TE guided-mode resonance.} \textbf{a}, Simulated transmittance map as a function of angle of incidence (AOI) along the $y$ direction for an electrode-free, ideal two-dimensional array with the unit cell defined in Fig.~1b. \textbf{b}, $\mathrm{Re}(E_x)$ field profiles in the $x$--$y$ plane at the resonance conditions marked as \textbf{i} and \textbf{ii} in \textbf{a}. The fields are evaluated at the mid-plane of the BTO layer. Dashed boxes indicate the outline of the TiO$_2$ nanopillar. Scale bars, 200 nm. \textbf{c}, Measured transmittance map as a function of AOI along the $y$ direction for the $2~\mathrm{mm} \times 2~\mathrm{mm}$ device. \textbf{d}, Transmittance spectra extracted from \textbf{c} at AOIs of $0^\circ$, $0.5^\circ$, and $1^\circ$.}
\label{fig:figS2}
\end{figure}

\begin{figure}
\centering
\includegraphics[width=0.9\linewidth]{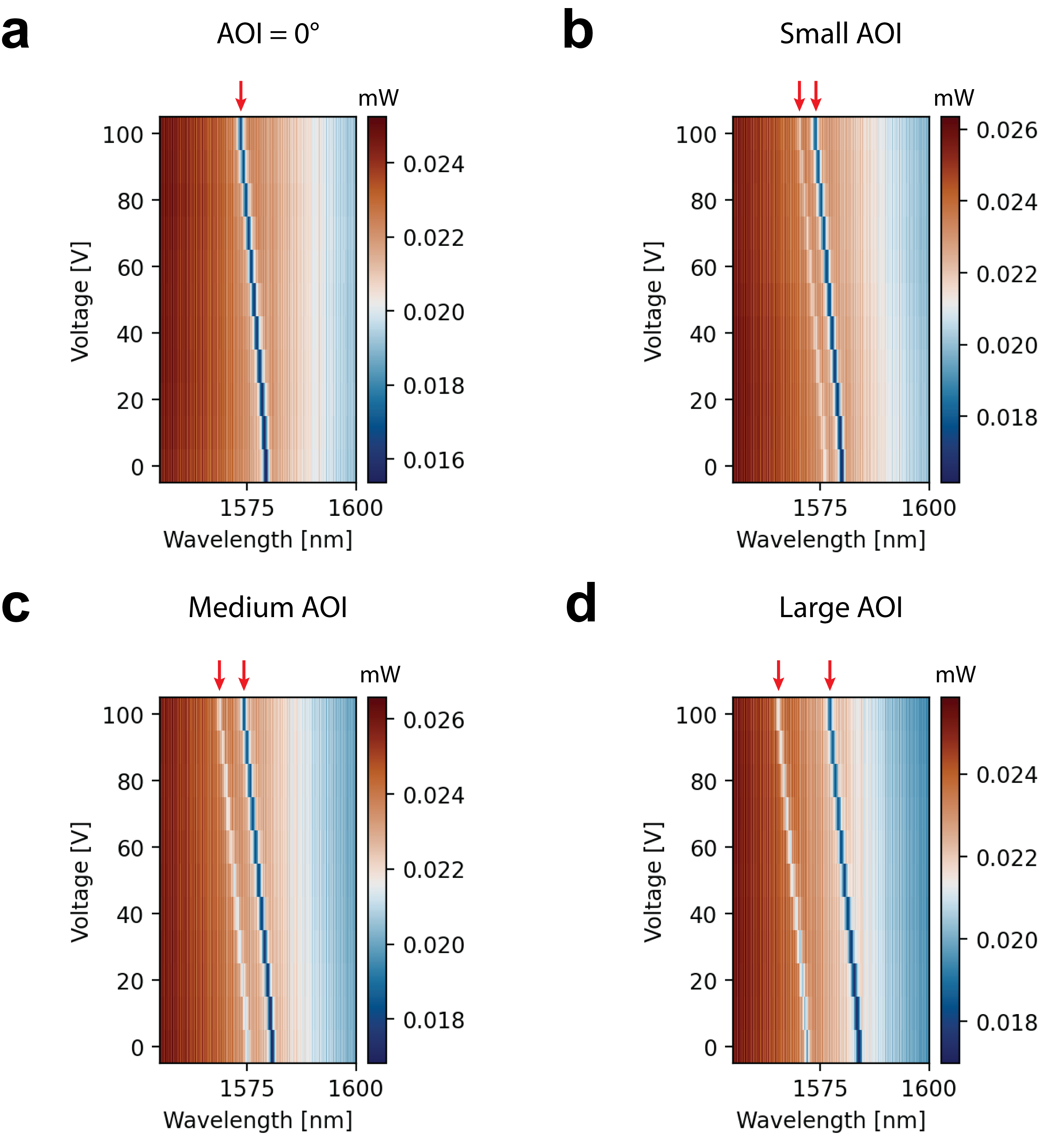}% Here is how to import EPS art
\caption{\textbf{Measured transmission maps as a function of wavelength and DC bias for the $2~\mathrm{mm} \times 2~\mathrm{mm}$ device at different angles of incidence (AOIs) along the $y$ direction.} The DC bias is swept from 0 to 100~V. \textbf{a}, Normal incidence. \textbf{b--d}, Three different oblique AOIs. The red arrows above the maps indicate the resonance positions. All maps show unnormalized raw transmission data.}

\label{fig:figS3}
\end{figure}

\begin{figure}
\centering
\includegraphics[width=1\linewidth]{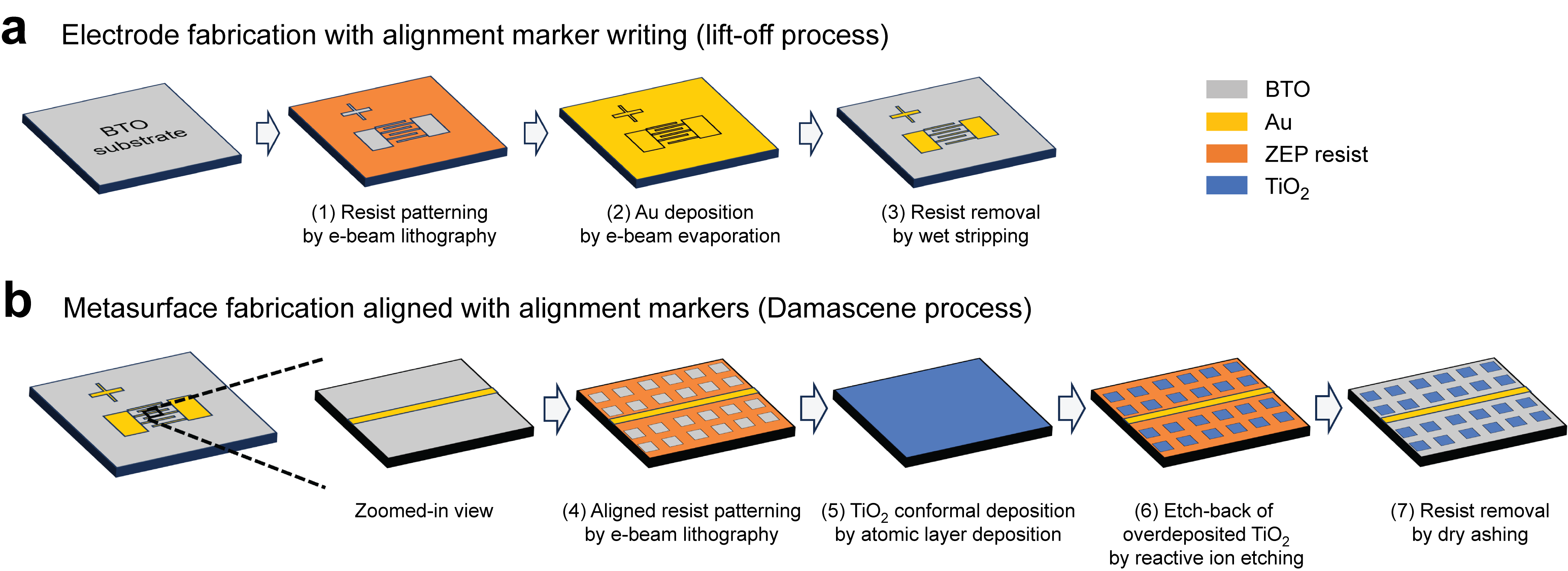}% Here is how to import EPS art
\caption{\textbf{Fabrication process for the hybrid BTO/TiO$_2$ metasurfaces.}}
\label{fig:figS4}
\end{figure}

\begin{figure}
\centering
\includegraphics[width=1\linewidth]{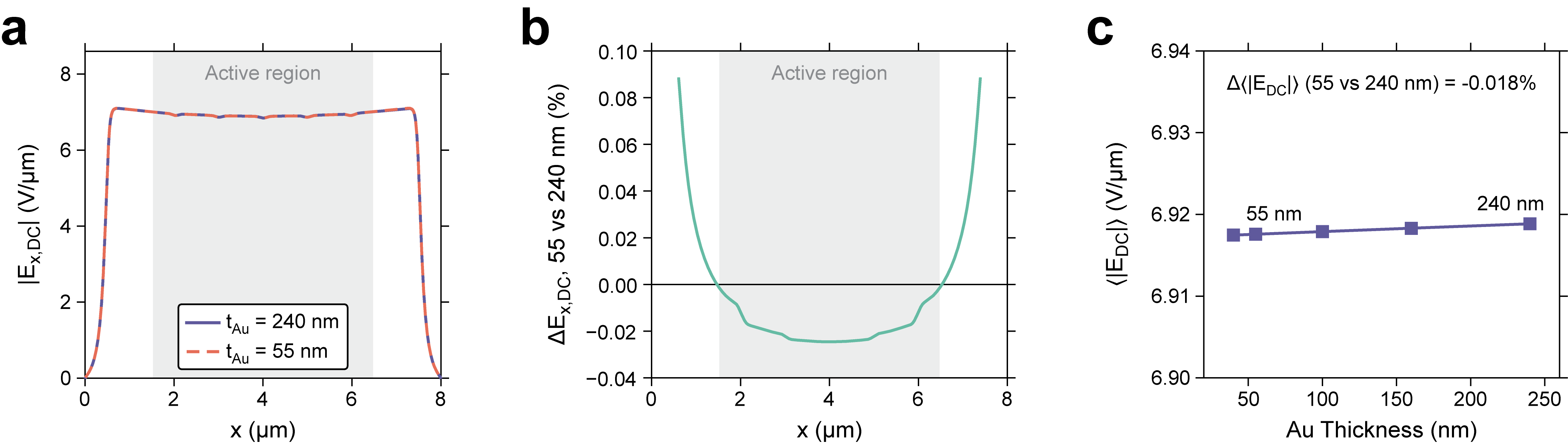}% Here is how to import EPS art
\caption{\textbf{Comparison of simulated static electric fields for different Au thicknesses $t_{\mathrm{Au}}$.} \textbf{a}, $E_x$ components of the static fields along the line of intersection between the $x$--$y$ mid-plane of the BTO layer and the $x$--$z$ plane through the centers of the TiO$_2$ nanopillars, for $t_{\mathrm{Au}} = 240$ and $55~\mathrm{nm}$. \textbf{b}, Relative difference in the $E_x$ field between $t_{\mathrm{Au}} = 55$ and $240~\mathrm{nm}$, normalized to the $E_x$ field for $t_{\mathrm{Au}} = 240~\mathrm{nm}$, along the same line as in \textbf{a}. \textbf{c}, Static field magnitude averaged over the active region, $\langle |E_{\mathrm{DC}}| \rangle$, as a function of $t_{\mathrm{Au}}$. The relative difference in $\langle |E_{\mathrm{DC}}| \rangle$ between $t_{\mathrm{Au}} = 55$ and $240~\mathrm{nm}$ is only $-0.018\%$.}
\label{fig:figS5}
\end{figure}

\begin{figure}
\centering
\includegraphics[width=1\linewidth]{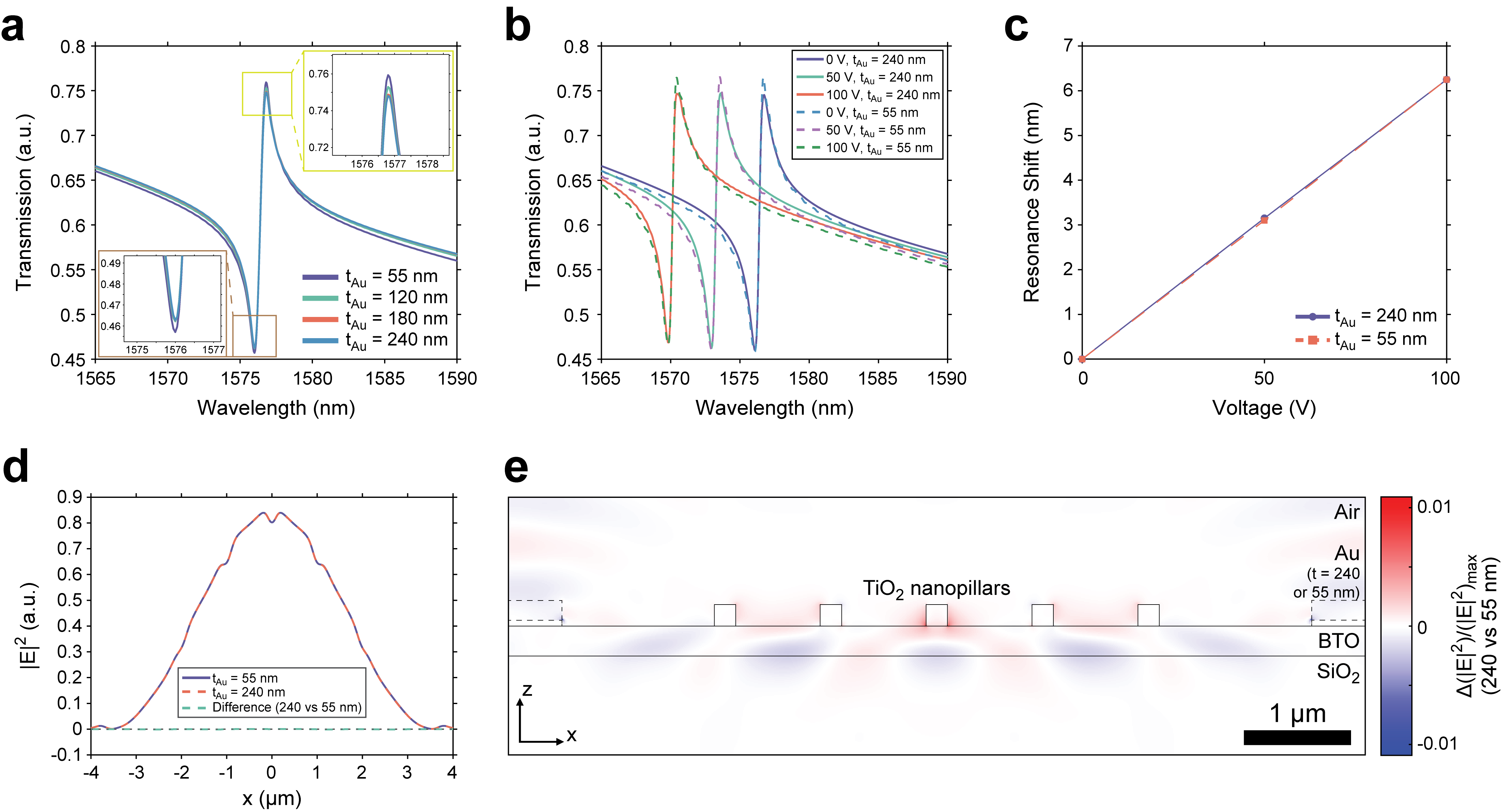}% Here is how to import EPS art
\caption{\textbf{Comparison of FDTD simulation results for supercells with different Au thicknesses $t_{\mathrm{Au}}$ at normal incidence.} \textbf{a}, Transmission spectra for different $t_{\mathrm{Au}}$. The insets show zoomed-in views of the Fano resonance peaks and dips. \textbf{b}, Transmission spectra for $t_{\mathrm{Au}} = 240~\mathrm{nm}$ (solid lines) and $55~\mathrm{nm}$ (dashed lines) under applied voltages of 0, 50, and 100~V. \textbf{c}, Resonance shift as a function of applied voltage for $t_{\mathrm{Au}} = 240$ and $55~\mathrm{nm}$, extracted from \textbf{b}. \textbf{d}, Squared electric-field magnitude, $|E|^2$, at the resonance along the line of intersection between the $x$--$y$ mid-plane of the BTO layer and the $x$--$z$ plane through the centers of the TiO$_2$ nanopillars, for $t_{\mathrm{Au}} = 240$ and $55~\mathrm{nm}$, together with their difference. \textbf{e}, Relative difference in $|E|^2$ at the resonance between $t_{\mathrm{Au}} = 240$ and $55~\mathrm{nm}$ in the $x$--$z$ plane through the centers of the TiO$_2$ nanopillars, normalized by the overall maximum $|E|^2$ value across the two fields in this plane.}
\label{fig:figS6}
\end{figure}

\begin{figure}
\centering
\includegraphics[width=0.8\linewidth]{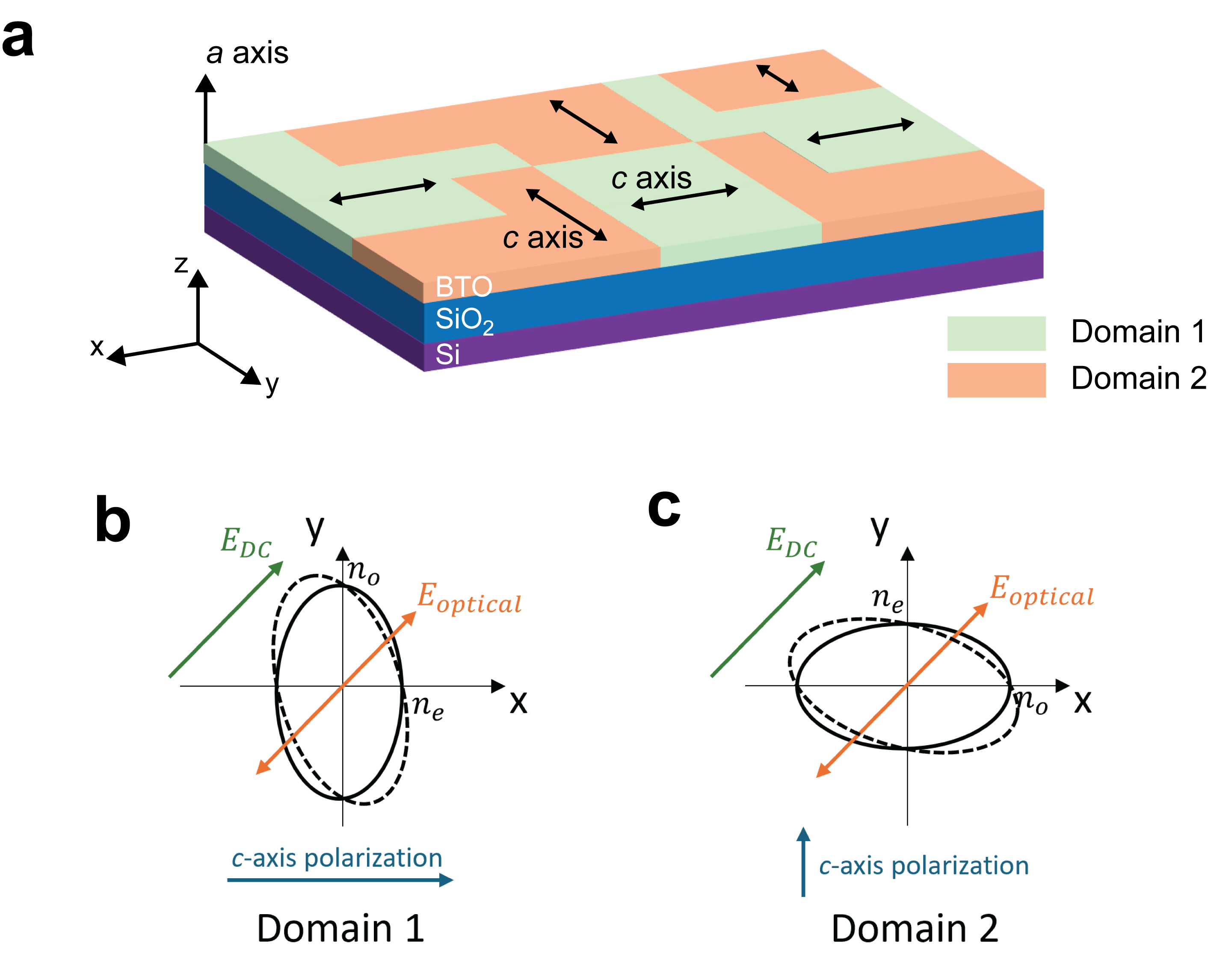}% Here is how to import EPS art
\caption{\textbf{$r_{42}$-mediated Pockels effect in orthogonal-domain $a$-oriented BTO.} \textbf{a}, Illustration of the two types of domains in thin-film orthogonal-domain $a$-oriented BTO. \textbf{b,c}, Refractive-index ellipsoids projected onto the $x$--$y$ plane for Domain~1 (\textbf{b}) and Domain~2 (\textbf{c}) of BTO. The static and optical electric fields are both oriented along $\frac{1}{\sqrt{2}}\left(\hat{\mathbf{x}} + \hat{\mathbf{y}}\right)$, and the ferroelectric polarization in each domain is aligned with the projection of the static field onto the $c$ axis of that domain. The unperturbed and static-field-perturbed refractive-index ellipses are shown as solid and dashed curves, respectively. The separation between the two ellipses along the optical-field direction represents the refractive-index change induced by the $r_{42}$-mediated Pockels effect.}
\label{fig:figS7}
\end{figure}

\begin{figure}
\centering
\includegraphics[width=1\linewidth]{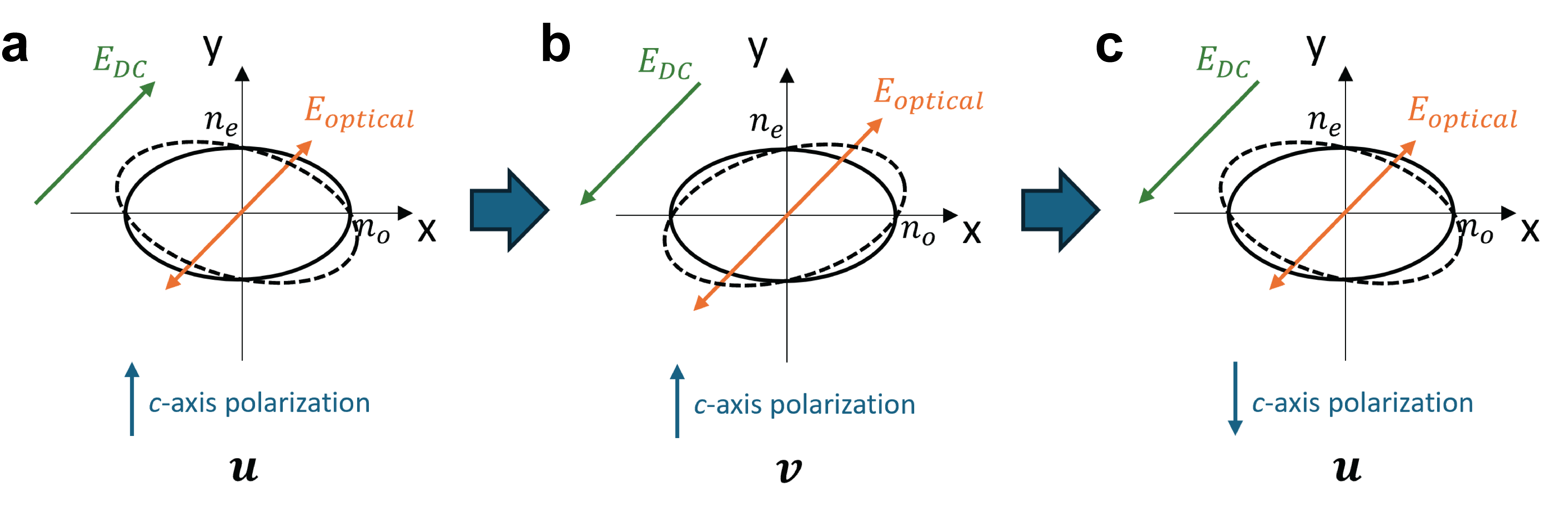}% Here is how to import EPS art
\caption{\textbf{Ferroelectric hysteresis in the $r_{42}$-mediated Pockels effect.} \textbf{a}, Initial state, taken from the configuration shown in Fig.~S7c. The refractive index for the indicated optical polarization is reduced relative to the unperturbed value. \textbf{b}, The static field reverses direction, while the ferroelectric polarization remains in its initial state due to hysteresis. The static-field-perturbed refractive-index ellipse (dashed) is inverted relative to that in \textbf{a}, corresponding to a sign reversal of the $r_{42}$-induced index perturbation. The refractive index for the indicated optical polarization is increased relative to the unperturbed value. \textbf{c}, The ferroelectric polarization reverses direction as the static-field strength increases. The static-field-perturbed refractive-index ellipse (dashed) returns to its initial configuration, corresponding to the recovery of the original sign of the $r_{42}$-induced index perturbation. The refractive index for the indicated optical polarization is again reduced relative to the unperturbed value. The configurations in \textbf{a} and \textbf{c} are defined as the $u$ state, whereas the configuration in \textbf{b} is defined as the $v$ state.}
\label{fig:figS8}
\end{figure}

\begin{figure}
\centering
\includegraphics[width=1\linewidth]{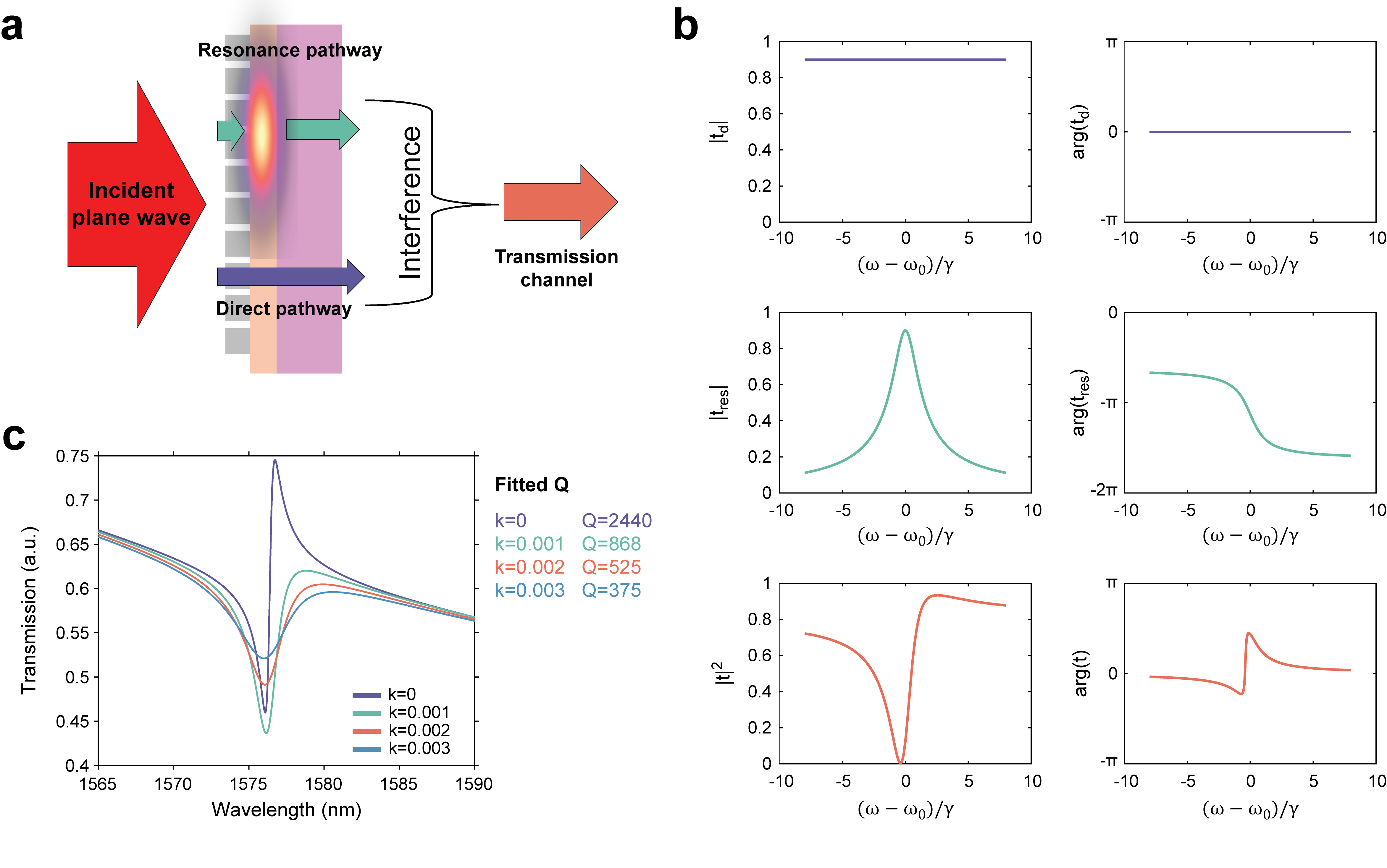}% Here is how to import EPS art
\caption{\textbf{Fano lineshape of the guided-mode resonance.} \textbf{a}, Schematic illustration of the origin of the Fano lineshape of the guided-mode resonance. \textbf{b}, Illustrative spectra of the magnitude and phase of the direct transmission amplitude $t_{\mathrm{d}}$ (top), the magnitude and phase of the resonant transmission amplitude $t_{\mathrm{res}}$ (middle), and the transmittance $|t|^2$ and phase of the total transmission amplitude $t=t_{\mathrm{d}}+t_{\mathrm{res}}$ (bottom). \textbf{c}, Simulated transmission spectra with different artificially introduced extinction coefficients $k$ in BTO and TiO$_2$. The $k=0$ curve is identical to the 0~V curve in Fig.~3d. The results show that the quality factor, transmission contrast, and Fano asymmetry all degrade with increasing $k$.}
\label{fig:figS9}
\end{figure}

\begin{figure}
\centering
\includegraphics[width=1\linewidth]{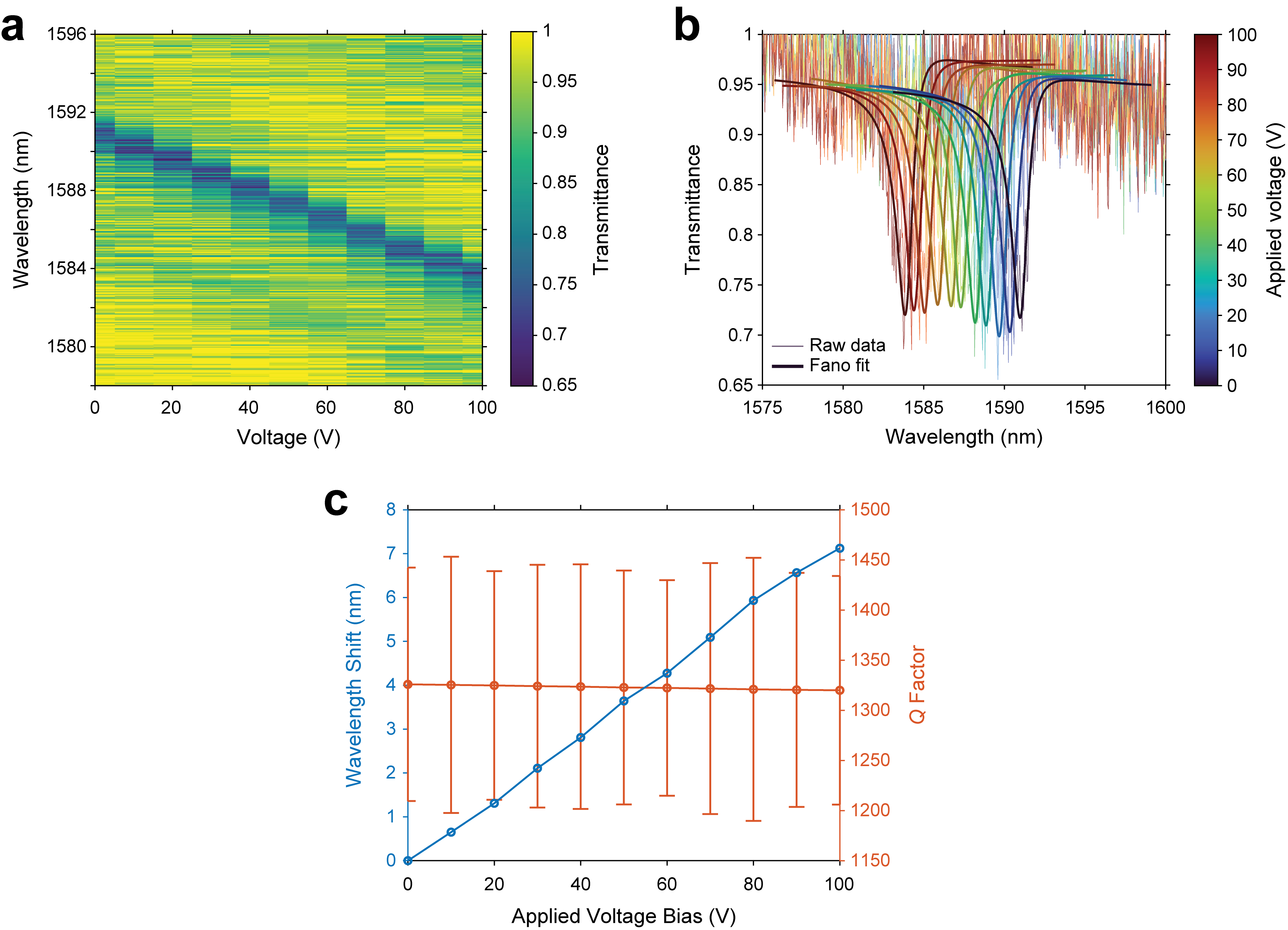}% Here is how to import EPS art
\caption{\textbf{Additional experimental results on DC modulation for the device with a $0.3~\mathrm{mm} \times 0.3~\mathrm{mm}$ metasurface.} \textbf{a}, Normalized transmittance map as the applied voltage is swept from 0 to 100~V. \textbf{b}, Normalized transmittance spectra (light, thin lines) and corresponding Fano fits (dark, thick lines) under different applied voltages. \textbf{c}, Resonance wavelength shift and quality factor extracted from the Fano fits as a function of the applied voltage. Error bars represent the standard errors from the Fano fits.}
\label{fig:figS10}
\end{figure}

\begin{figure}
\centering
\includegraphics[width=1\linewidth]{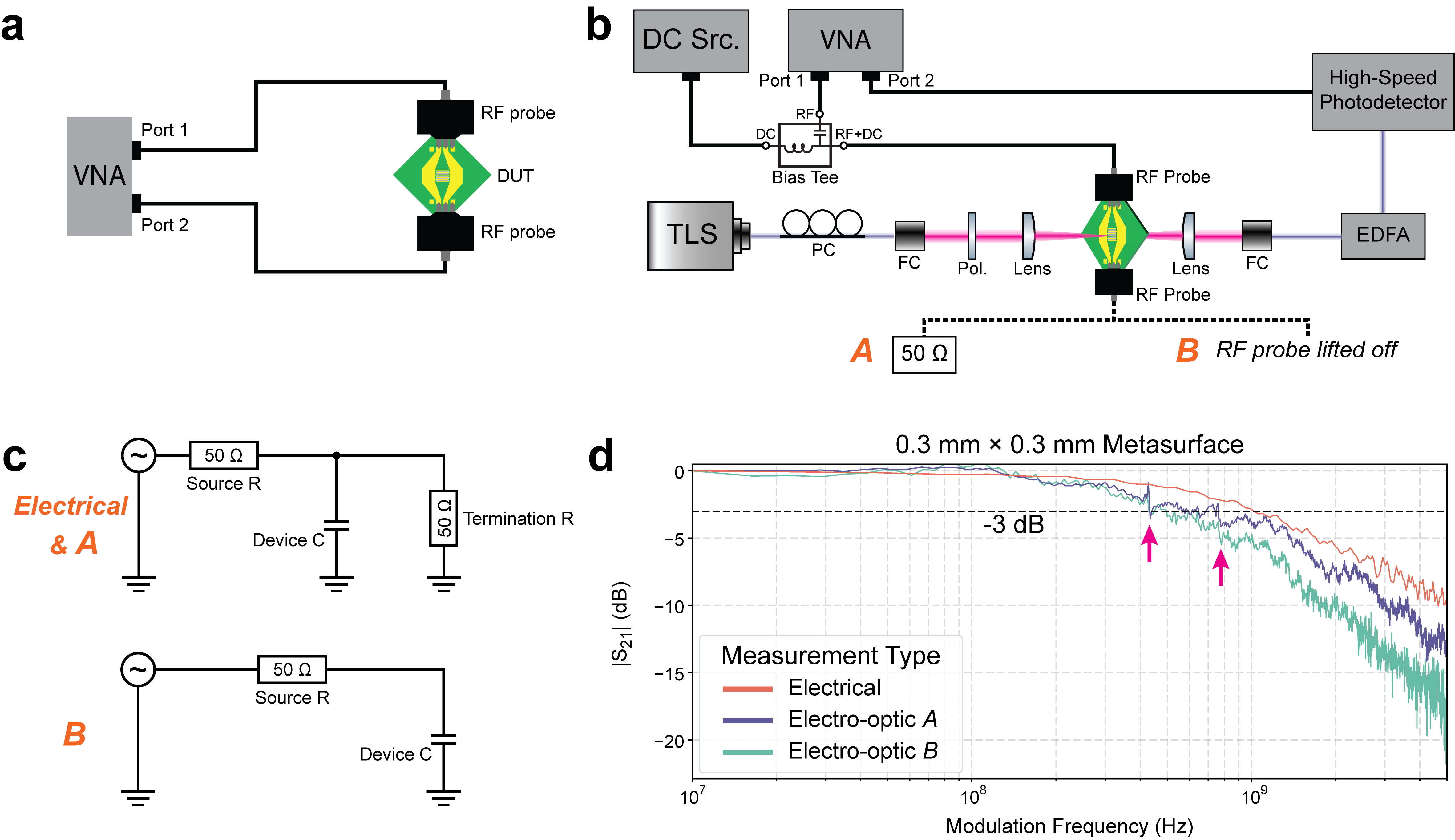}% Here is how to import EPS art
\caption{\textbf{Additional experimental results on high-speed modulation for the device with a $0.3~\mathrm{mm} \times 0.3~\mathrm{mm}$ metasurface.} \textbf{a}, Measurement setup for electrical bandwidth characterization. \textbf{b}, Measurement setup for electro-optic bandwidth characterization. Two RF termination configurations, \textit{A} and \textit{B}, are studied. Configuration \textit{A} is identical to that in Fig.~4b, with a 50~$\Omega$ resistor connected to the bottom RF probe, corresponding to a terminated RF circuit. In configuration \textit{B}, the bottom RF probe is lifted off from the contact pads, corresponding to an unterminated RF circuit. VNA, vector network analyzer; DUT, device under test; DC Src., DC voltage source; TLS, tunable laser source; PC, polarization controller; FC, fiber collimator; Pol., polarizer; EDFA, erbium-doped fiber amplifier. \textbf{c}, Equivalent circuit models for the electrical configuration and EO configuration \textit{A} (top), and EO configuration \textit{B} (bottom). Only the RF equivalent circuits are shown; the DC-bias branch of the bias tee is omitted for clarity. \textbf{d}, High-speed modulation responses for the electrical configuration and EO configurations \textit{A} and \textit{B}. The resonance-like signatures marked by arrows in the responses for configurations \textit{A} and \textit{B} suggest the presence of acoustic resonances.}
\label{fig:figS11}
\end{figure}

\FloatBarrier
\bibliographystyle{naturemag}
\bibliography{BTOSIref}